\documentclass[manuscript,screen]{acmart}

\usepackage{amsmath,amsfonts}
\usepackage{algorithmic}
\usepackage{algorithm}
\usepackage[caption=false,font=normalsize,labelfont=sf,textfont=sf]{subfig}
\usepackage{textcomp}
\usepackage{stfloats}
\usepackage{url}
\usepackage{verbatim}
\usepackage{graphicx}
\usepackage{multirow} 

\AtBeginDocument{%
  }

\setcopyright{acmlicensed}
\copyrightyear{2018}
\acmYear{2018}
\acmDOI{XXXXXXX.XXXXXXX}
\acmConference[Conference acronym 'XX]{Make sure to enter the correct
  conference title from your rights confirmation email}{June 03--05,
  2018}{Woodstock, NY}
\acmISBN{978-1-4503-XXXX-X/2018/06}

\begin{document}

\title{Selecting and Combining Large Language Models for Scalable Code Clone Detection}

\author{Muslim Chochlov}
\affiliation{%
  \institution{University of Limerick}
  \department{Department of Computer Science and Information Systems}
  \city{Limerick}
  \country{Ireland}
}
\email{muslim.chochlov@ul.ie}

\author{Gul Aftab Ahmed}
\affiliation{%
  \institution{Trinity College Dublin}
  \department{Department of Computer Science}
  \city{Dublin}
  \country{Ireland}
}
\email{ahmedga@tcd.ie}

\author{James Vincent Patten}
\affiliation{%
  \institution{University of Limerick}
  \department{Department of Computer Science and Information Systems}
  \city{Limerick}
  \country{Ireland}
}
\email{james.patten@lero.ie}

\author{Yuanhua Han}
\affiliation{%
  \institution{Huawei Technologies Co., Ltd.}
  \department{WN Digital IPD and Trustworthiness Enabling}
  \city{Xi'an}
  \state{Shaanxi}
  \country{China}
}
\email{hanyuanhua2@huawei.com}

\author{Guoxian Lu}
\affiliation{%
  \institution{Huawei Technologies Co., Ltd.}
  \department{WN Digital IPD and Trustworthiness Enabling}
  \city{Shanghai}
  \country{China}
}
\email{luguoxian@huawei.com}

\author{David Gregg}
\affiliation{%
  \institution{Trinity College Dublin}
  \department{Department of Computer Science}
  \city{Dublin}
  \country{Ireland}
}
\email{david.gregg@tcd.ie}

\author{Jim Buckley}
\affiliation{%
  \institution{University of Limerick}
  \department{Department of Computer Science and Information Systems}
  \city{Limerick}
  \country{Ireland}
}
\email{jim.buckley@ul.ie}

\renewcommand{\shortauthors}{Trovato et al.}

\begin{abstract}
Source code clones pose risks ranging from intellectual property violations to unintended vulnerabilities. Effective and efficient scalable clone detection, especially for diverged clones, remains challenging. Large language models (LLMs) have recently been applied to clone detection tasks. However, the rapid emergence of LLMs raises questions about optimal model selection and potential LLM-ensemble efficacy.

This paper addresses the first question by identifying 76 LLMs and filtering them down to suitable candidates for large-scale clone detection. The  candidates were evaluated on two public industrial datasets, BigCloneBench, and a commercial large-scale dataset. No uniformly 'best-LLM' emerged, though CodeT5+110M, CuBERT and SPTCode were top-performers. Analysis of LLM-candidates suggested that smaller embedding sizes, smaller tokenizer vocabularies and tailored datasets are advantageous. On commercial large-scale dataset a top-performing CodeT5+110M achieved 39.71\% precision: twice the precision of previously used CodeBERT.

To address the second question, this paper explores ensembling of the selected LLMs: effort-effective approach to improving effectiveness. Results suggest the importance of score normalization and favoring ensembling methods like maximum or sum over averaging. Also, findings indicate that ensembling approach can be statistically significant and effective on larger datasets: the best-performing ensemble achieved even higher precision of 46.91\% over individual LLM on the commercial large-scale code.
\end{abstract}





\maketitle

\section{Introduction}

Code clones are sections of code that are duplicated within or across projects and are similar to a certain extent \cite{Rattan2013, Bellon2007}. Developers might copy and paste (duplicate) code to save time, reuse code templates, or unknowingly implement similar functionalities \cite{Rattan2013, Ain2019}. While this practice can speed up development initially, it often leads to problems with software maintenance, vulnerability detection (e.g., when updates or fixes aren't applied uniformly across all the clones, potentially introducing bugs or security issues), and intellectual property violation (e.g. when copyright disallows copying or is not appropriately acknowledged)\cite{chochlov2022using}.

Code clones are commonly classified according to their textual similarity. For example, a frequently used classification framework divides clones into these four types \cite{Bellon2007, Rattan2013, Ain2019, Svajlenko2014a}:
\begin{enumerate}
    \item Type 1 clones are identical pieces of code, but allowing for different comments and white-spacing;
    \item Type 2 clones additionally allow for identifier and variable renaming;
    \item Type 3 clones additionally allow for insertion, deletion, or modification of execution statements;
    \item Type 4 clones bear little textual similarity yet are functionally identical.
\end{enumerate}

Svajlenko et al. propose an even finer classification for Type 3 clones (again using their textual similarity) dividing the category further into very strong Type 3 (VST3: 90\% - 100\% textually similar), strong Type 3 (ST3: 70\% - 90\% textually similar), and moderate Type 3 (MT3: 50\% - 70\% textually similar) \cite{Svajlenko2017a}. They also introduced a weak category of Type 3 / Type 4 (WT3 / T4) for clones that are less than 50\% textually similar \cite{Svajlenko2017a}.

As software projects grow larger and increasingly complex, finding these code clones manually becomes challenging and impractical: instead automated approaches are needed that are both \textit{effective} (can accurately locate clones) and \textit{efficient} (e.g. time-efficient) \cite{Rahman2020, chochlov2022using, ahmed2024nearest}. Many traditional non AI-based \cite{Kamiya2002654, Li2020, Jiang2007, Sajnani2016, Hung2020, Wu2020} and AI-based \cite{wang2020detecting, alon2019code2vec, saini2018oreo} clone detection techniques (CDTs) have been introduced over the years \cite{Bellon2007, Rattan2013, Sheneamer2016, Ain2019}. These CDTs are usually characterized according to the source code information that is extracted and used for clone detection, resulting in text-based, token-based, abstract syntax tree (AST)/program dependency graph (PDG) based, and metrics-based categories, along with an associated hybrid category \cite{Rattan2013}. Some of these CDTs are effective and efficient when locating Type 1/2 clones, but their effectiveness starts to decrease when locating Type 3 clones \cite{wang2020detecting, saini2018oreo}. 

% (As was the experience of our industrial partner when using a token-based, scalable CDT, CCFinder \cite{Kamiya2002654}: they required support for Type 3 clone detection, using more appropriate techniques \cite{chochlov2022using, ahmed2024nearest}).

More recently, Vaswani et al. \cite{vaswani2017attention} introduced a novel generation of artificial neural networks (ANN) with the \textit{transformer} architecture that can be applied to the task. These ANNs, with hundreds of millions/billions of parameters, are frequently referred to as Large Language Models (LLMs). Following their success in natural language tasks \cite{Devlin2018, radford2018improving}, LLMs have subsequently made significant advances in the understanding-of and generation-of programming language(s) \cite{xu2022systematic, Wong2023, dou2023towards}, enabling tasks such as code completion, code synthesis, error detection, and clone detection \cite{Devlin2018}. Early attempts at applying LLMs to clone detection particularly, showed that these LLMs are effective \cite{feng2020codebert, guo2020graphcodebert}, but that their pairwise-comparison approach is unsuitable for large codebases. For example, early work in this area \cite{feng2020codebert} tasked the CodeBERT LLM with classification of each possible two segments of code as clones or not, but \(N\) source code segments then requires \(O (N^2)\) comparisons, which is impractical for any large, real-world application (where \(N\)might refer to the number of individual methods in the application or even every-possible 7-line segment in the application).

Instead, in our previous work, an LLM-agnostic approach called SSCD was introduced that leverages the ability of LLMs to generate\textit{ code embeddings} (numerical representations of code). Coupled with k-approximate nearest neighbour (kANN) search, this can effectively and efficiently locate clones at scale \cite{chochlov2022using, ahmed2024nearest}, even those clones that have diverged somewhat: Empirical evaluation of this approach showed state-of-the-art results in clone detection \cite{chochlov2022using}.

%Using LLMs for code clone detection offers a promising avenue. Because they can understand the underlying meaning of code, LLMs have the potential to identify clones that traditional methods might miss, especially those that look different syntactically but perform the same function. This semantic understanding could lead to more accurate and scalable clone detection solutions.

Yet despite their potential for large-scale clone detection, LLMs have not been fully leveraged in this role. One of the reasons for this is the large number of LLMs continuously appearing \cite{minaee2024large}, and thus the difficulty for researchers in making informed decisions on the LLMs to apply. To the best of our knowledge, there are no studies systematically exploring LLM characteristics such as size, training data or architecture in the clone-detection context, and so there is currently little understanding of how those characteristics impact on large-scale clone detection. In addition, the comparative studies that do exist are few-and-far between \cite{Khajezade2024, Niu2023}, do not focus on large-scale clone detection and/or are limited in terms of the number of LLMs they assess: For example, Khajezade et al. only look at the ChatGPT LLMs \cite{Khajezade2024}.

Another issue is variability in individual LLM performance, suggesting that reliance on a single model may not be optimal. Ensemble inference (i.e., combining the result-sets of multiple LLMs) has shown promise in other domains for improving robustness and effectiveness \cite{Sagi2018}. It is an under explored area in clone detection \cite{ahmed2023using} and questions remain as to how best to construct and optimize ensembles effectively.

This work aims to address these issues by identifying unique novel LLMs, suitable for scalable code clone detection and evaluating them for that task. Six specific characteristics of these LLMs are studied to understand how they correlate with recall/precision performance, providing guidance to researchers and practitioners on selecting the appropriate ones for clone detection. Recognizing that no single model may be perfect, ensemble inference of the top-performing models is assessed to determine whether their combined strengths can lead to better performance. Therefore, this work answers the following research questions (RQs):

\begin{itemize} 
\item RQ1. How effective are novel LLMs for scalable clone detection? 
\item RQ2. How do characteristics of these LLMs affect their effectiveness with regard to recall? \item RQ3. How effective are ensembles of these LLMs? 
\begin{itemize}
    \item RQ3a. How do ensembling methods affect effectiveness?
    \end{itemize}
\end{itemize}

This work makes the following contributions to the field of large-scale code clone detection:
\begin{enumerate}
     
\item We identify and assess LLMs for scalable code clone detection, filtering 76 unique models from prior literature, through structured criteria, to an evaluation that identifies CodeT5+ 110M, CuBERT and SPTCode as top performers.
\item We provide significant empirical evidence that  model performance can be highly dataset dependent, LLM-architecture dependent, training dependent and tokenizer-vocabulary dependent. In terms of dataset-dependency, for example, we show that models like ``CodeT5+ 110 M embedding'' \cite{wang2023codet5+} excel on datasets with smaller clone classes, while others like CodeT5 \cite{Wang2021} and StarEncoder\cite{li2023starcoder} perform better on benchmarks with larger clone classes. The evidence provided also suggests that larger embedding sizes, multilingual training, and larger tokenizer vocabulary negatively impact recall, while training on datasets beyond CodeSearchNet improves performance. We further demonstrate that scaling model parameters alone does not enhance recall, aligning with recent trends that emphasize data quality over model size. These insights offer actionable guidance for optimizing LLM selection for code clone detection.

\item We introduce a Borda count aggregation method to provide a holistic ranking, ensuring fair model comparison across datasets and across models, extending our LLM-selection methodology, and recognising CuBERT \cite{kanade2020learning} to be the top-performing, most stable model for clone detection.

\item  From our trialling of LLM ensembles we have found evidence that ensembling multiple LLMs offers statistically significant improvements on \emph{larger} or more diverse codebases (e.g., BigCloneBench), but often seems to underperform on \emph{smaller} datasets. This nuanced finding suggests that the efficacy of ensembling depends on dataset size and diversity, rather than guaranteeing universal gains.

\item We demonstrate that \emph{specific} normalizations (e.g., \texttt{min-max}, \texttt{z-score}, \texttt{rrf}) and aggregations (\texttt{sum}, \texttt{max}) are vital to achieving ensemble benefits, and suggest best practice in this context. For (counter) example, the \texttt{non-norm/average} approach systematically reduces efficacy below even the levels of single models, highlighting what \emph{not} to do when combining LLM result-sets for large-scale clone detection.

\item  Through an evaluation of LLMs on a private industrial dataset, we illustrate that real-world clone detection outcomes diverge from public benchmark results, but that ensembling does result in an efficacy gain: Despite its mixed performance across public datasets, ``CodeT5+ 110 M embedding'' \cite{wang2023codet5+} achieves the highest precision (39.71\%) and detects significantly more true clones (552) than alternative models. Ensembling pushes precision even higher. While the best absolute precision improvement was from 39.71\% to 46.91\%, the highest relative improvement was 37.43\%, obtained when precision increased from 18.06\% to 24.82\%. This highlights the critical role of real-world evaluation in assessing LLM efficacy beyond standard benchmarks. 
\end{enumerate}

%To summarize, By addressing the challenges of scalable code clone detection through innovative applications of LLMs and ensemble methods, we aim to develop more efficient and effective tools for software maintenance. Our findings have practical implications for software engineers and researchers looking to manage and evolve large codebases more effectively, reducing the risk of errors and increasing productivity.
The paper is organized as follows: Section~\ref{sec:background} provides background knowledge of transformer-based LLMs, gives a brief overview of SCCD, provides an overview of LLM-ensembling strategies and discusses related work in assessment of LLMs towards clone detection. In Section~\ref{sec:methodology}, the research questions are revisited, the process of identification/filtration of novel LLMs for Clone Detection is described, and the experimental methodology is presented. In Section~\ref{sec:results} assessment results are presented followed by discussion of these results. In section~\ref{sec:threats} threats to the validity of the experiments are presented. Finally, section~\ref{sec:conclusions} summarizes this work and discusses future work directions.   

\section{Background and Related Work}
\label{sec:background}
\subsection{Transformer-based LLMs, their architectures, and characteristics}
\label{subsec:arch_char}
Transformer-based LLMs are modern ANNs originally proposed by Vaswani et al. \cite{vaswani2017attention}. These ANNs couple \textit{self-attention} mechanism and \textit{parallelization} allowing for learning complex data dependencies in large datasets \cite{vaswani2017attention}. The ANNs of this type subsequently were proven to be effective in a variety of natural language tasks and programming language tasks\cite{Devlin2018, radford2018improving, feng2020codebert, guo2020graphcodebert}.

Such LLMs rely on several main components, particularly, the tokenizer, the encoder and the decoder \cite{vaswani2017attention}. For programming language tasks, the tokenizer processes input sequences, such as source code, and converts the input into a set of unique numerical identifiers that can be understood by an LLM, using the tokenizer's vocabulary \cite{kudo2018sentencepiece, sennrich2015neural, schuster2012japanese}. The encoder then processes these numerical identifiers, and generates rich, contextualized representations of the data (e.g., code embeddings). It does this through a series of layers (where each layer consists of a number of parameters), each of which applies multi-head self-attention and feed-forward neural networks. The decoder, on the other hand, is tasked with generating output sequences based on the intermediate representations (code embeddings) provided by the encoder. LLMs can be encoder-only models (e.g., BERT\cite{Devlin2018}, RoBERTa\cite{liu2019roberta}), decoder-only models (e.g., auto-regressive models like GPT\cite{brown2020language}), and encoder-decoder models (e.g., T5\cite{raffel2020exploring}, BART\cite{lewis2019bart}). In LLMs designed for tasks like clone detection, the decoder may be omitted entirely, as the goal could be to produce code-classification embeddings rather than generate sequences (of new code, for example).

LLMs need to be trained before they can be used meaningfully. In terms of training, transformer-based LLMs typically undergo two major stages in their development: pretraining and fine-tuning \cite{Devlin2018, feng2020codebert}. Pretraining involves training the model on large, generic datasets (e.g., in case of code related LLMs, these datasets might be programming language repositories such as Github) using unsupervised learning objectives such as masked language modeling or next-token prediction\cite{Devlin2018}. This pretraining step allows the model to learn general-purpose representations of the input data. Fine-tuning adapts the pretrained model to specific downstream tasks, such as code clone detection, using task-specific labeled datasets\cite{guo2020graphcodebert}. For example, the CodeBERT LLM \cite{feng2020codebert} is pretrained on CodeSearchNet \cite{husain2019codesearchnet}, a large dataset of over 6 million functions written in 6 programming languages, and later fine-tuned on a smaller BigCloneBench-derived \cite{Svajlenko2017a} dataset for clone detection, specifically. 

Transformer-based LLMs can be described using a set of characteristics stemming from their architectures and training. To the best of our knowledge there is currently no finite set of LLM characteristics and therefore the set of characteristics used in this work is derived from prior work in the field (including our previous work). For example, Wong et al. \cite{Wong2023} and Shervin et al. \cite{minaee2024large} highlight common characteristics such as training datasets and supported programming languages. Meanwhile, our previous work with language models \cite{chochlov2022using, ahmed2024nearest} has emphasized characteristics related to tokenization and inference length. While other research has highlighted the importance of the number of parameters \cite{Touvron2023}. Combined together these characteristics can be presented as follows:

\begin{itemize}
    \item Architectural Characteristics
    \begin{itemize}
        \item Type of architecture family: encoder-only, decoder-only, or encoder-decoder models influence the scope and focus of the LLM.
        \item Number of layers: can be used as a proxy for the depth of the model, often represented by the total number of parameters.
        \item Number of parameters: can determine the model's capacity to learn and generalize complex patterns.
        \item Embedding size: represents the dimensionality of the representations used internally by the model.
    \end{itemize}
    \item Training-Related Characteristics
    \begin{itemize}
        \item Data size: The amount of data used during pretraining, often serving as a proxy for the diversity and scale of knowledge encoded in the model.
        \item Dataset composition: refers to the nature of the data used for pretraining (e.g., general web text, programming language repositories).
        \item Programming languages: for code related LLMs, the languages included in the pretraining corpus, which will likely affect the model's ability to handle specific programming tasks in specific programming languages.
    \end{itemize}
    \item Inference-Related Characteristics
    \begin{itemize}
        \item Context length: refers to the maximum length of the input sequence that the model can process.
    \end{itemize}
    \item Tokenizer: the method used to preprocess input sequences into tokens. Tokenizers can have multiple sub-parameters and, to avoid over-saturation, can be evaluated holistically.
\end{itemize}

\subsection{Illustrative example: code tokenization and embedding with CodeBERT LLM}
\label{subsec:il_example}
\begin{figure}[htbp]
\includegraphics[width=0.5\textwidth]{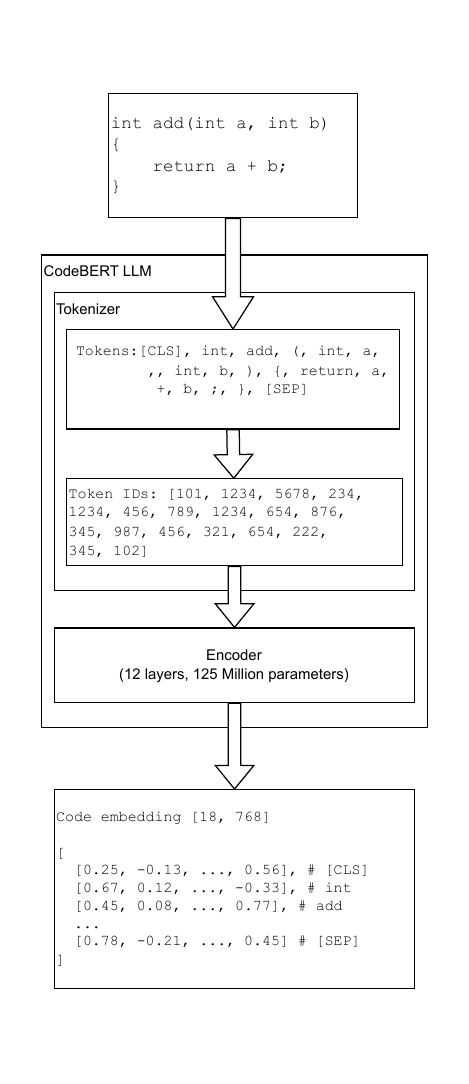}
\caption{Illustrative example: code inference with CodeBERT LLM}
\label{fig:example}
\end{figure}

To demonstrate how a tokenizer and encoder work together, a CodeBERT \cite{feng2020codebert} model with 12 layers and 125 million parameters is used to generate code embeddings for a simple C function as shown in Figure~\ref{fig:example}. The process begins with the tokenizer, which breaks the input code into smaller units, referred to as tokens. Tokens can represent keywords, identifiers, operators, or even structural elements such as parentheses or braces. In CodeBERT, special tokens are also added to the sequence to guide the model's processing. For instance, the `[CLS]` token is placed at the beginning of the sequence, serving as a representation of the entire input, while the `[SEP]` token marks the end of the sequence or separates multiple segments. After this initial tokenization step, the above C function is transformed into the sequence of 18 tokens, as shown in Figure~\ref{fig:example}. Next, each token is mapped to a unique token ID using CodeBERT’s predefined vocabulary. These IDs are numerical representations that allow the model to process the tokens computationally. For example, the token `int` is mapped to the ID `1234`, while `add` corresponds to `5678`. Special tokens like `[CLS]` and `[SEP]` are also assigned specific IDs, such as `101` and `102`, respectively.

These token IDs are then passed to the encoder, which processes the sequence to generate contextualized embeddings for each token. For CodeBERT (base variant), the size of these embeddings is 768 dimensions. Therefore, the output of the encoder is a matrix of embeddings \(E(18, 768)\), where each row corresponds to one of the 18, identified tokens, and each column represents one of the 768 dimensions. For example, the `[CLS]` token might be represented as: \([0.25, -0.13, ..., 0.56]\).

\subsection{SSCD overview}
\label{subsec:sscd}
%related work
\begin{figure*}[htbp]
\centerline{\includegraphics[width=\textwidth]{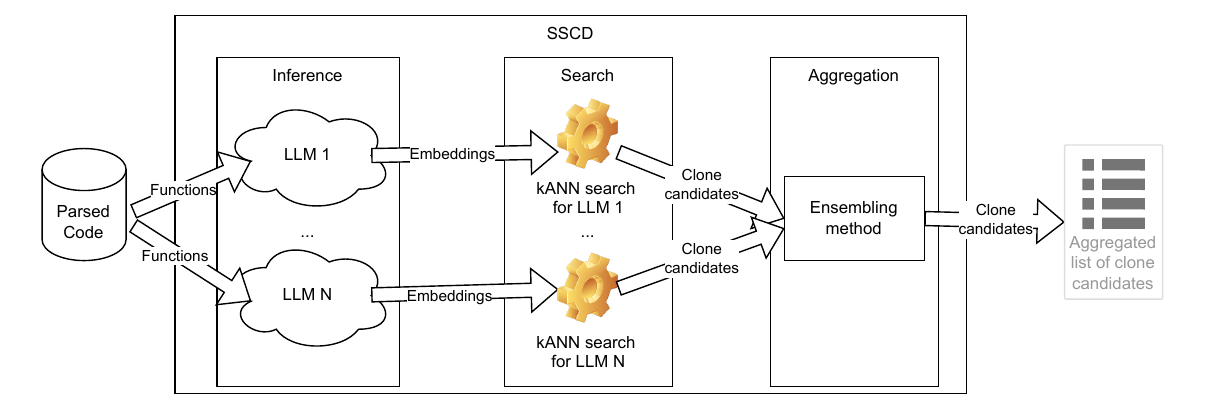}}
\caption{Diagram of SSCD}
\label{fig:sscd}
\end{figure*}
In this work, SSCD and its LLM-agnostic (the ability to accommodate heterogeneous transformer-based LLMs) capabilities are used as a harness for the experiments. Here, a brief description of SSCD is provided as a foundation for our work: for a detailed description of SSCD our previous papers should be consulted \cite{chochlov2022using, ahmed2023using, ahmed2024nearest}. 

As can be seen in Figure~\ref{fig:sscd}, the SSCD has three major components. First, the parsed code (a set of functions) is feeded into an `inference` module. In this module, one or more\footnote{In this work we used two-model ensembles, although there is no limitation on the number of LLMs to be used.} LLMs can be used to generate code embeddings: one code embedding per function, per LLM. The inference here follows steps described in Section~\ref{subsec:il_example}. Adding to these steps, an average of all token embeddings is taken, similar to Reimers \cite{reimers2019sentence} to allow for a single vector representation per function: Using the illustrative example from Section~\ref{subsec:il_example}, the code embedding \(E(18, 768)\) is reduced to \(E (1, 768)\). In the next, `search` module, generated code embeddings are compared to each other using k approximate nearest neighbor (kANN) algorithms \cite{Malkov2018, Johnson2019} to find the most similar embeddings, efficiently. Embedding pairs are ranked globally according to their similarity and mapped to actual functions to produce ranked lists of clone candidates. Finally, in cases where an ensembled version of clone detection is desired, the `aggregation` module, merges the results of various LLMs (lists of clone candidates) together using a user-selected ensembling method. The final list of aggregated clone candidates has the following format: let the list of clone candidates be denoted as \( L \), where:
\[
L = \{(p_1, s_1), (p_2, s_2), \dots, (p_n, s_n)\}
\]
Here:
\begin{itemize}
    \item \( p_i \) represents a pair of functions \( (e_{i1}, e_{i2}) \).
    \item \( s_i \) represents the similarity score of the pair \( p_i \).
\end{itemize}

For each function \( e_{ij} \) (\( j = 1, 2 \)) in the pair \( p_i \):
\[
e_{ij} = (\text{path}_{ij}, \text{start}_{ij}, \text{end}_{ij})
\]
where:
\begin{itemize}
    \item \( \text{path}_{ij} \): The file path for the function.
    \item \( \text{start}_{ij} \): The start line in the file.
    \item \( \text{end}_{ij} \): The end line in the file.
\end{itemize} 

For example, given C language functions over six files, and cosine similarity, an SSCD-generated list of the potential clone candidates could look as follows:
\[
L = \left\{
\begin{aligned}
  &(((\text{fileA.c}, 10, 20), (\text{fileB.c}, 15, 25)), 0.85), \\
  &(((\text{fileC.c}, 5, 15), (\text{fileD.c}, 8, 18)), 0.80), \\
  &(((\text{fileE.c}, 1, 5), (\text{fileF.c}, 3, 7)), 0.75)
\end{aligned}
\right\}
\]

In practice, SSCD commonly returns \textit{clone classes}: sets of clone candidates for one piece of code. Re-using and modifying the example above, if (\text{fileA.c}, 10, 20) had three similar candidates then these are called its clone class and the size of this class is the number of its candidates, i.e. 3:
\[
L = \left\{
\begin{aligned}
  &(((\text{fileA.c}, 10, 20), (\text{fileB.c}, 15, 25)), 0.85), \\
  &(((\text{fileA.c}, 10, 20, (\text{fileD.c}, 63, 71)), 0.74), \\
  &(((\text{fileA.c}, 10, 20, (\text{fileF.c}, 20, 31)), 0.72)
\end{aligned}
\right\}
\]

The set of SSCD parameters, that are essential for understanding evaluations presented in this paper is as follows:
\begin{itemize}
    \item \texttt{code\_length} is the maximum number of tokens that are used as an LLM input. If a code piece has more tokens than this maximum number, then these are truncated at \texttt{code\_length}.
    \item \texttt{minloc} is the minimum number of LOC for a piece of code: pieces of code below this number are discarded by SSCD.
    \item \texttt{similarity threshold} is the minimum similarity between clone candidates in a clone pair/class: candidates below this threshold are discarded.
    \item \texttt{top N clone class candidates} is the maximum number of clone candidates in a clone class returned.
    \item \texttt{global top K} is the maximum number of all clone candidates returned, based on the \texttt{K} nearest-neighbors in the embedding
\end{itemize}

\subsection{LLM ensembling strategies}
Ensembling is combining multiple models in an attempt to improve the overall effectiveness for a downstream task, by leveraging the strengths of the constituent models \cite{ahmed2023using, Sagi2018}. With transformer-based LLMs, ensembling has been applied to a variety of domains, including classification, regression, and sequence modeling \cite{ghosh2021using}.

Conceptually, ensembling of LLMs can be divided into the following strategies \cite{wan2024knowledge}:
\begin{itemize}
    \item Aggregation of results: the final outputs of LLMs are combined. This method allows for ensembling architecturally-heterogeneous LLMs at the expense of resource-efficiency: the models have to be executed in parallel (affecting memory resources mostly) or sequentially (affecting total execution time).
    \item Merging of models' architectures and/or their parameters, to achieve better results. This method might require architecturally homogeneous models and is non-trivial to implement\cite{wan2024knowledge}.
    \item Stacking ensembles, where the outputs of multiple base models are fed into a meta-model that learns how to optimally combine their predictions. For example, in multi-agent LLM setups, stacking has been employed to coordinate predictions from multiple specialized models, resulting in improved accuracy and task-specific alignment \cite{zhangmore}. It has also been trialed in our preliminary work on ensembles, but didn't prove to be effective \cite{ahmed2023using}.
\end{itemize}

In this work, the first method (aggregation of results) is employed, because of the heterogeneous nature of the LLMs reviewed. Specific result-aggregation methods in this category can be distinguished here:

\begin{itemize}
    \item Voting-based ensembles, where predictions from multiple models are aggregated to form a consensus opinion. For instance, majority voting is commonly used for classification tasks, while weighted voting can account for varying confidence levels of the models involved \cite{Sagi2018}. In the context of LLMs, such approaches have been successfully applied to tasks like scientific article categorization \cite{ghosh2021using}.

    \item Model averaging, where the probabilistic outputs (e.g., softmax distributions) of multiple models are merged to generate final predictions. This strategy is particularly effective for tasks that require robust uncertainty quantification. For LLMs, such as those used for knowledge fusion, combining the outputs of different models can effectively aggregate specialized knowledge domains, leading to superior performance in tasks requiring broad contextual understanding \cite{wan2024knowledge}.
\end{itemize}

Here, SSCD returns lists of clone candidates (see Section~\ref{subsec:sscd}) where two pieces of code in a clone candidate are measured by two heterogeneously-derived similarity scores. This makes the 'Model averaging" score-fusion approach seem more suitable. Several score-fusion methods can be applied here: taking the maximum score of either model, summarizing the scores (e.g. if two models agree on a clone candidate this will promote it up the ranking list), and averaging the scores to gain a more balanced score. We explore several of these methods in this work.

%Additionally, **task-specific ensembling strategies** are emerging as powerful solutions for complex applications. In the context of large-scale LLMs, ensembling has been applied to tasks like knowledge fusion, where the strengths of multiple LLMs are combined to produce cohesive and accurate outputs (Wan et al., 2024). These strategies often integrate task-specific heuristics, such as dynamic weighting based on model confidence or context sensitivity.

\subsection{Assessment of LLMs towards clone detection}
Despite the large number of LLMs appearing in recent years \cite{minaee2024large, xu2022systematic}, their assessment for large-scale clone detection has been limited.

Dou et al. \cite{dou2023towards} explored the performance of Large Language Models (LLMs) in detecting code clones across various programming languages. Particularly, they used two LLMs, CodeBERT and text-embedding-ada-002 (OpenAI) to generate code embeddings for clone detection and assessed the two models on the BigCloneBench dataset. Their findings indicated that text-embedding-ada-002 was better, achieving a higher F-score. But in their work, the number of LLMs was limited to two, the LLM characteristics related to efficacy in the task were not studied and LLM-ensembling was not assessed.

Khajezade et al. \cite{Khajezade2024} studied the effectiveness of the ChatGPT (OpenAI) model for clone detection using pairwise clone classification. The model was compared to three other LLM - RoBERTa, CodeBERT, and GraphCodeBERT - and showed an improvement in terms of F-score on these models. Unlike this work, the authors did not focus on scalable clone detection (assessing pairwise code classification), their selection of LLMs was limited to 4, LLM characteristics were not studied, and ensembling was not assessed.

Niu et al. \cite{Niu2023} trialed 19 pre-trained LLMs for a variety of software engineering tasks including clone detection. For clone detection, a subset of LLMs was selected, including PLBART, CodeT5, and SynCoBERT. These models were used for pairwise clone detection (non-scalable) using the BigCloneBench and CLCDSA datasets. The authors also tried to categorize LLMs, but based on their suitability for software engineering tasks. Again this is different from current work, where the focus is on scalable clone detection leveraging LLMs, studying their architectural characteristics towards improved performance, and assessing their ensembles.

%This information is then passed to either fingerprinting, vectorization, information retrieval (IR) and suffix-tree algorithms for clone detection \cite{Rattan2013, Ain2019}. For example, SourcererCC is a scalable token-based CDT utilizing IR for clone detection \cite{Sajnani2016}.

%Some token-based CDTs utilizing suffix-trees like CCFinder \cite{Kamiya2002654} and SAGA \cite{Li2020} 
 
%Likewise, as discussed above, another approach commonly seen in CDTs is to transform code into a numerical representation: for example, a vector in a vector-space, a hash-based fingerprint, or a metrics-based vector. These representations are then used with clustering, search, or IR algorithms to find similar representations \cite{Jiang2007, Sajnani2016, Hung2020, Wu2020}. The latter approaches still lack in their clone detection effectiveness, efficiency, or both \cite{Li2020, chochlov2022using}. 
%Artificial intelligence was used in CDTs with mixed results, particularly in terms of their efficiency \cite{wang2020detecting, alon2019code2vec, saini2018oreo}: for example, Oreo was estimated to execute for 222 hours on a 250 MLOC clone benchmark \cite{Li2020}, finding all possible clones (batch clone search). For comparison, the execution time for SAGA was reported as 1 hour and 24 minutes for the same task and dataset \cite{Li2020}.  

\section{Methodology}
\label{sec:methodology}
\subsection{Research objectives}
The research objectives of this work are as follows:

\begin{itemize}
    \item To evaluate the effectiveness of newer LLMs for scalable code clone detection in terms of their efficacy. Here, this work aims to systematically identify and assess the performance of transformer-based LLMs in detecting code clones at scale (addressing RQ1).
    \item Using the evaluations, to analyze the impact of LLM characteristics on clone detection performance in terms of its recall. The goal here is to investigate how specific characteristics of LLMs, such as architecture type, number of parameters, training datasets, supported programming languages, and tokenization strategies (see Section~\ref{subsec:arch_char}), influence their performance in code clone detection tasks (addressing RQ2).
    \item Following the analysis of those characteristics, to suggest a framework for selecting LLMs for scalable clone detection. A resultant model is proposed to guide the selection of LLMs based on their characteristics, offering researchers and practitioners a predictive framework for identifying the most suitable models for scalable code clone detection (leading on from the findings for RQ2).
    \item To study ensembling of LLMs to improve clone detection. The goal here is to explore if and how ensembling, can enhance the effectiveness and robustness of LLMs in detecting code clones by combining the strengths of individual models (addressing RQ3 and RQ3a).
    \item Evaluate the applicability of top-performing LLMs and their ensembles in an industrial setting. This work, in particular, will evaluate the real-world performance of scalable LLM-based clone detection approaches, providing preliminary ecological evaluation for our findings.
\end{itemize}

\subsection{Identification of LLMs for scalable clone detection}

\begin{figure}[htbp]
\centerline{\includegraphics[width=0.5\textwidth]{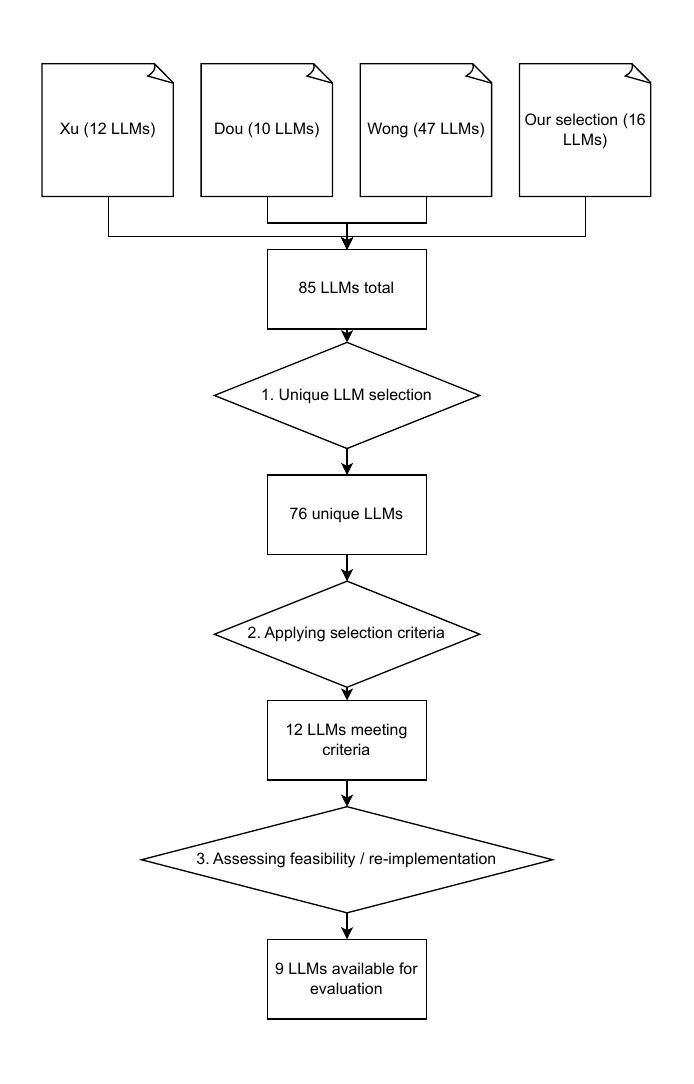}}
\caption{LLM selection process}
\label{fig:selection}
\end{figure}
In terms of identifying relevant LLMs, we focused on existing articles reviewing code-related LLMs, employing the following search protocols for data sources, search strategy, and inclusion criteria:

Data sources:
\begin{itemize}
    \item Peer-reviewed journals and conference proceedings;
    \item Preprints and arXiv papers.
\end{itemize}

Search Strategy:
\begin{itemize}
    \item Using Google Scholar as a search aggregator for scientific databases like Scopus, IEEE Xplore, and others due to its inclusive nature, where pre-prints from technical groups (proposing new LLMs) would also be included.
    \item Using keywords like "language model", "code", "software" and boolean operators: e.g. ("language model") AND ("code" OR "software") AND ("study" OR "evaluation").
\end{itemize}

Inclusion Criteria:
\begin{itemize}
    \item Papers published within the last 5 years (2019-2024);
    \item Studies focusing on LLM for software engineering approaches (particularly, clone detection);
    \item Systematic literature reviews and other meta-analyses.
\end{itemize}

Using this approach we identified three reviews: Xu et al.\cite{xu2022systematic}, Dou et al.\cite{dou2023towards}, and Wong et al\cite{Wong2023}.

The schema in Figure~\ref{fig:selection} then, outlines the process of selecting LLMs for scalable code clone detection. Initially, LLMs were collected by aggregating the LLMs from the three reviews identified.  This includes contributions from Xu et al. \cite{xu2022systematic} (the 12 LLMs selected, can be found in Figure 1 of their paper), Dou et al. \cite{dou2023towards} (the 10 LLMs selected, can be found in Table 2), and Wong et al. \cite{Wong2023} (the 47 LLMs selected, can be found in Table 4). But these reviews were published in 2022, 2023 and 2023, meaning that the LLMs reviewed were possibly the state-of-the-art in 2021, 2022 and 2022 respectively. Given the speed with which new LLMs are released, we augmented this initial set of LLMs, employing our own expertise of code-related LLMs, based on our clone-detection work to date. Our expertise is evidenced by our four peer-reviewed publications on this matter published in top tier venues \cite{chochlov2022using, ahmed2024nearest, ahmed2023using, gul2025}. A further 16 LLMs were thus selected for review. These are: Code Llama \cite{roziere2023code}, SantaCoder \cite{allal2023santacoder}, CodeGen2 \cite{nijkamp2023codegen2}, StarEncoder \cite{li2023starcoder}, StarChat Alpha \cite{Tunstall2023starchat-alpha}, Replit Code \cite{replit_code_v1_3b}, CodeT5+ 110M embedding \cite{wang2023codet5+}, CodeT5+ 220M \cite{wang2023codet5+}, Xgen-7B \cite{nijkamp2023xgen}, Codegen2.5 \cite{nijkamp2023codegen2}, Decicoder-1B \cite{DeciFoundationModels}, Palm 2 \cite{anil2023palm}, Chinchilla \cite{hoffmann2022training}, Lamda \cite{thoppilan2022lamda}, Megatron-Turing NLG \cite{smith2022using}, and Gopher \cite{rae2021scaling}. 

Altogether, this process resulted in a combined pool of 85 LLMs. Here, different versions of the same LLM, or different variants of the same LLM, were considered separate models. For example, GPT-2 and GPT3.5 were considered separate models; likewise, Llama 7B \cite{Touvron2023}(the model with 7 billion parameters) and LLama 27B \cite{Touvron2023} (the model with 27 billion parameters). 

The following four steps were then used to select the final set of LLMs for evaluation:
\begin{itemize}
    \item Redundant LLMs, across the different reviews, were eliminated resulting in 76 unique LLMs.
    \item Only LLMs pre-trained (and/or fine-tuned) for source code tasks were retained. This ensured that the models have a foundational understanding of the syntactic and semantic structures of code.
    \item Public availability: to ensure reproducibility and accessibility for both academic and industrial applications, only publicly available LLMs were considered. Proprietary models or restricted-access systems were excluded.
    \item Support for embeddings (having an encoder as part of their architecture): the selected LLMs must support the generation of embeddings through an encoder-based architecture. This criterion ensures that the models can effectively encode source code functions into meaningful vector representations suitable for similarity/clone detection tasks.
\end{itemize}

Tools that merely utilized LLMs but did not provide direct access to the underlying model or embeddings were also excluded. Following this step, the pool of LLMs was reduced down to 12 remaining models. 

The final step of the selection process involved assessing the practical feasibility of using or re-implementing (if necessary) the selected models. Of the 12 models selected 8 were available and were integrated into SSCD, 1 model (C4 \cite{tao2022c4}), was fine-tuned according to the documentation and integrated into SSCD, and 3 models could not be re-implemented and/or integrated. Regarding the latter 3 models, TreeGEN \cite{sun2020treegen} and Recoder \cite{zhu2021syntax} did not have a C/C++ tokenizer available and re-implementing such a tokenizer would be a significant undertaking in its own right; TBCC could not be trained \cite{hua2022transformer}. The final set of 9 LLMs were: 
\begin{enumerate}
    \item CodeBERT (fine-tuned) \cite{feng2020codebert, chochlov2022using} (CBFT);
    \item CodeT5 \cite{Wang2021} (CT5);
    \item GraphCodeBERT \cite{guo2020graphcodebert} (GCB);
    \item CuBERT \cite{kanade2020learning} (CuBERT);
    \item StarEncoder \cite{li2023starcoder} (StarEncoder);
    \item SPT-Code \cite{niu2022spt} (SPTCode);
    \item C4 \cite{tao2022c4} (C4);
    \item CodeT5+ 110M embedding \cite{wang2023codet5+} (CT5P-110);
    \item CodeT5+ 220M \cite{wang2023codet5+} (CT5P-220).
\end{enumerate}

\subsection{Common Evaluation framework}
Common core components of an experimental design are presented in this section. These include datasets, metrics, and hardware configurations.

\subsubsection{Datasets}
\begin{table}[!th]
\caption{Clone datasets used for evaluation}
\label{tbl:datasets}
\centering
\resizebox{0.5\textwidth}{!}{\begin{tabular}{lrrlr}
\toprule
\textbf{Dataset}        & \textbf{Total LOC} & \textbf{Availability} & \textbf{Language} & \begin{tabular}[c]{@{}r@{}}\textbf{\# Clone Pairs} \\ \textbf{@ Method-Level}\end{tabular} \\ 
\midrule
Company-C               & 61 KLOC            & Public                & C                 & 77                                     \\ 
Company-C++             & 315 KLOC           & Public                & C++               & 85                                     \\ 
BCB13                   & 13 MLOC            & Public                & Java              & 8,375,313                              \\ 
In-situ-C-C++           & 13/300 MLOC            & Private               & C, C++            & N/A                                    \\ 
\bottomrule
\end{tabular}}
\end{table}

The evaluations in this paper were carried out using four data sets, as shown in Table~\ref{tbl:datasets}. These datasets represent a set of three programming languages (C/C++/Java) and range from small to large code sizes.
\begin{itemize}
\item Company-C and C++ datasets were provided to us by out commercial partners and they have 61 KLOC (thousand lines of code) and 315 KLOC, of C and C++ code respectively. These datasets are publicly available \cite{SSCD_Dataset_01} and include 77 C method-level clone pairs and 85 C++ method-level clone pairs. Both datasets were created by company experts to represent real-world C/C++ clone cases encountered by the company.
\item The BigCloneBench (BCB13) \cite{Svajlenko2017a} dataset is a large-scale, public \cite{BigCloneBench} dataset consisting of 13 MLOC (million lines of code) in Java in its reduced version. It includes an extensive set of 8,375,313 clone pairs at the method level, making it a standard benchmark for assessing scalability and performance in substantial code repositories \cite{Sajnani2016, chochlov2022using}, despite recent criticism of its quality \cite{krinke2022bigclonebench}.
\item  Finally, the in-Situ-C-C++ dataset, containing 13 MLOC across four C and four C++ systems, represents a private codebase developed internally by our industrial partner. It is compared against a 300 MLOC open source public codebase for clone detection purposes. Due to its in-situ nature, the absolute number of method-level clone pairs in this dataset is unknown, impacting the metrics employed in associated evaluations. However, its in-situ nature enables real-world evaluation and allows us to assess the practical utility of the proposed approaches.
\end{itemize}

\subsubsection{Metrics}
To assess the effectiveness of clone detection using various LLMs with SSCD, two key metrics were used: \textbf{recall} and \textbf{precision}. These are often reported in combination to provide a comprehensive evaluation of a CDT's performance \cite{Rattan2013}. Recall quantifies how many known clones a CDT successfully identifies, while precision evaluates the accuracy of the clones detected by an approach.

Mathematically, these metrics are defined as follows:
\[
\text{Recall (\%)} = \frac{\text{True Positives}}{\text{True Positives} + \text{False Negatives}} * 100
\]
\[
\text{Precision (\%)} = \frac{\text{True Positives}}{\text{True Positives} + \text{False Positives}} * 100
\]

For example, if a CDT detects 25 true clones out of 100 known clones, the recall is:
\[
\text{Recall} = \frac{25}{25 + 75} * 100 = 25\%
\]
Likewise, if the CDT returns 100 clone candidates, of which 30 are correct, the precision is:
\[
\text{Precision} = \frac{30}{30 + 70} * 100 = 30\%
\]

For Company-C and C++ datasets (see Table \ref{tbl:datasets}), both recall and precision can be calculated: the definitive set of clones in these datasets is known to the company's experts, who created them \cite{chochlov2022using}. For BCB13, only recall of 'known' clones can be calculated automatically \cite{Svajlenko2017a}, because the absolute number of clones in BCB13 is unknown. Likewise precision, even though some studies evaluate precision by taking a sample of the clone candidates returned by a Clone Detection approach for assessment by a human evaluator: such an approach can be subject to bias \cite{Li2020}. Finally, the In-Situ-C-C++ dataset is a real-world codebase and the amount of clones in that codebase is unknown. To calculate precision, a system expert has to examine a list of returned clone candidates, and this also tells us the (relative) numbers of clones identified by an approach.

\subsubsection{Hardware configuration}
In this work, two machines were used for evaluation: An M1 machine with i7-10875H 2.3GHz 8 core CPU, 32 GB RAM, Quadro T2000 4GB GPU, 1TB SSD (evaluation of public datasets) and a C1 machine with 16 GB RAM, Tesla T4 16 GB GPU (in-situ evaluation at the company).

\subsection{Experimental design for RQ1}
\label{subsec:rq1_design}
To answer RQ1, we trialed the 9 identified novel LLMs in a scalable manner using 4 datasets (see Table~\ref{tbl:datasets}) and assessed their effectiveness. Three high level steps can be distinguished as part of the experimental design employed:

\begin{enumerate}
    \item We assessed 9 selected LLMs on three publicly available datasets (Company-C, Company-C++, and BCB13);
    \item We ranked these LLMs according to their effectiveness, aggregating these ranks across these three public datasets;
    \item We then selected the best-performing LLMs and assessed these LLMs in-situ at the company using the company's private dataset In-Situ-C-C++.
\end{enumerate}

For step 1, we executed the 9 LLMs in a scalable manner (using SSCD's nearest-neighbour approach), for the three datasets (Company-C, Company-C++, BCB13) and the search was at function-level granularity. The selection of SSCD parameters when assessing Company-C and Company-C++ was as follows:

\begin{enumerate}
    \item \textbf{Constant parameters}: \texttt{code\_length}, \texttt{minloc}, \texttt{similarity threshold}, and \texttt{top N clone class candidates} were constant and set to 128 (\texttt{code\_length}), 0 (\texttt{minloc}), 0 (\texttt{similarity threshold}), and 10 (\texttt{top N clone class candidates}). We set \texttt{code\_length} to 128 tokens based on our previous findings, where increasing the size of this parameter resulted in only small effectiveness gains, but decreased efficiency significantly \cite{chochlov2022using}. We set \texttt{minloc} to 0 to include all functions in the datasets irrespective of their length. We set \texttt{similarity threshold} to 0 because LLMs commonly return distinct similarity score distributions, and setting this value arbitrarily might disadvantage an LLM in an ensemble: setting this to 0 allows an LLM to return all possible clone candidates found. Finally, we set \texttt{top N clone class candidates} to 10, knowing that clone class sizes are equal to 1 in these two datasets but allowing for more flexibility and again larger sets of clone classes for more representative evaluation.
    \item \textbf{Variable parameters}: \texttt{global top K} was set for Company-C++ to [10, 50, 83, 166] and for Company-C to [10, 50, 70, 140]. Essentially, this allows for a more nuanced effectiveness evaluation with a very small number of clone candidates returned (10), a slightly larger set at 50, a set of clone candidates the same size as the number of true clones in each dataset (83 for Company-C++ and 70 for Company-C), and a set of clone candidates twice the size of true clones (166 and 140 respectively).
\end{enumerate}

For BCB13, all parameters were constant. Similarly to Company-C and Company-C++ trials, we set \texttt{code\_length} to 128, again based on our previous findings \cite{chochlov2022using}. We set \texttt{minloc} to 10 following best practices for evaluation of this dataset \cite{chochlov2022using, Li2020, Svajlenko2017a}. We set \texttt{top N clone class candidates} to 100, knowing that BCB13 has large clone classes \cite{chochlov2022using}. \texttt{Global top K} was set to 3,000,000, which is approximately 10\% of all possible code pairs that could potentially be returned: at \texttt{minloc}=10, there are 306,290 functions in BCB13 and, allowing for each such function to have a potential clone candidate class of size 100, results in 30,629,000 total clone candidates returned. We have not trialed other \texttt{global top K} parameters, knowing from our previous work with this dataset \cite{chochlov2022using} that this threshold (3,000,000) provides comparable representative results. Finally, \texttt{similarity threshold} was set differently for different LLMs to limit their initial output, due to the very large number of potential clone candidates but always surpassing the \texttt{top K} specified above.

With these parameters in place, on M1 machine we obtained 36 execution results for Company-C and Company-C++ datasets each (9 models $\times$ 4 \texttt{global top K} values) and 9 execution results for BCB13 (9 models $\times$ 1 \texttt{global top K}).  Following this, we calculated recall because maximizing recall was a core objective of our study. Also, recall is a common metric that can be calculated for all three datasets. In contrast, calculating precision for BCB13 is not trivial \cite{Svajlenko2017a} and often requires human evaluation that can be biased \cite{Li2020, krinke2022bigclonebench}. For Company-C and Company-C++, recall is calculated at each of four \texttt{global top K} values for each dataset, resulting in \texttt{Recall@10}, \texttt{Recall@50}, \texttt{Recall@83}, and \texttt{Recall@166} for Company-C++, and \texttt{Recall@10}, \texttt{Recall@50}, \texttt{Recall@70}, and \texttt{Recall@140} for Company-C. The average recall is then calculated as:
\[
\text{Average of recalls} = \frac{\sum \texttt{Recall@N}}{\texttt{Number of Recall Values}}
\]

For BCB13, BigCloneEval benchmark \cite{Svajlenko2017a} returns recall per clone type (T1, T2, VST3, ST3, MT3, and WT3/T4). Here the average recall is calculated as:

\[
\text{Average recall} = \frac{\sum \texttt{Recall@Type}}{\texttt{Number of Recall Types}}
\]

For step 2, we ranked all models according to their average recall. To compare the models across all three datasets and to find the best-performing models, we adopted Borda count \cite{van2000variants} based on their rankings. The Borda count is calculated as follows: 
\[
Borda Count = \texttt{N} + 1 - \texttt{ranking}
\]

where N is the number of participants. For example, if model A is ranked 5 for the Company-C dataset, then its Borda count is equal to 5 ($9 + 1 - 5$). Then all Borda counts are summed, and the final list of models is ranked according to these summed scores. We also looked at the standard deviation of rankings. For example, a model can rank 1 for two datasets but then rank 7 for another dataset, making this model's recall less stable. Conversely, another model can show more consistent results, ranking 2, 2, and 3 across the datasets, for example.

For step 3, we selected the best-performing LLMs for in-situ evaluation at the company with the In-Situ-C-C++ dataset. Due to the company's operational constraints, we could select 2 such LLMs. The following criteria were looked at when selecting these LLMs:

\begin{enumerate}
    \item \textbf{The rank of these LLMs}: the LLMs appearing in top positions according to their final Borda scores seem to be natural candidates due to their high effectiveness across all datasets.
    \item \textbf{Predictability of these LLMs}: those of the top LLMs that show more stable rankings (looking at standard deviation) across all datasets can be preferable.
    \item \textbf{Uniqueness of these LLMs' results}: those of the top LLMs belonging to different architecture families (see Section~\ref{subsec:arch_char}), for example, can be more advantageous when later trialing their ensembles, due to their perceived ability to identify unique clone candidates. Here, we looked at the (maximum) symmetric difference of their BCB13 results because these results are much larger than Company-C and Company-C++, and therefore can suggest a better representation of uniqueness. The symmetric difference is calculated as:
\[     A \Delta B = (A \cup B) - (A \cap B)     \] A higher symmetric difference signifies more distinct sets of clone candidates and therefore can suggest more distinct LLMs.
\end{enumerate}

Following this, the selected LLMs were executed on the company's C1 machine using their private In-Situ-C-C++ dataset for evaluation. Similar to previous trials, the execution was conducted at function-level granularity and in a scalable batch search manner: the private codebase of 13 MLOC was compared against the public codebase of 300 MLOC. The parameters were set as follows:
\begin{itemize}
    \item \texttt{code\_length} was set to 512 with the intent to maximize effectiveness, even if just marginally \cite{chochlov2022using}.
    \item \texttt{minloc} was set to 6 reflecting the company's policy on the minimal function's size suitable for clone detection.
    \item \texttt{top N clone class candidates} was set to 1, reflecting very small clone classes: essentially meaning that a function can have only one clone de3tected, in line with the company's software engineers' intuition.
    \item \texttt{similarity threshold} was set differently for trialled models. Conducting a batch search of 13 MLOC codebase against a 300 MLOC codebase can return a large number of clone candidates even with clone classes of size 1: selecting similarity threshold cut-off can reduce these further. Similarity thresholds per model were selected empirically and such they maximized the F-scores of the models when trialed on Company-C and Company-C++ datasets. 
    \item \texttt{global top K} was set to 1390. This number was not selected arbitrarily but to compare trials in this work with our previous trials on the same dataset with the CBFT model \cite{chochlov2022using}. In that previous trial, a system expert inspected the list of clone candidates and marked them as clones or not clones, as long as the precision@N (where N is the number of clone candidates inspected) was above or equal 20\%. In other words, a system expert would continue as long as at least 1-in-5 clone candidates is a true positive. To objectively compare new LLMs with CBFT the same \texttt{global top K} number was used.
\end{itemize}
With these parameters, the company experts collected execution results for top-performing models and measured the number of true-positives identified and precision. (Recall here cannot be measured because the number of clones in the dataset is unknown.) Particularly, a company expert inspected clone candidates and decided if they were clones or not-clones, discarding any duplicates. 

Combined, these trials provided evidence as to theeffectiveness of the 9 novel LLMs for scalable clone detection.

\subsection{Experimental design for RQ2}
To answer RQ2, we studied how characteristics (features) of the 9 models analysed in this work can affect their clone detection recall. The objective directing this question was hybrid:
\begin{itemize}
    \item To investigate how individual characteristics affect recall;
    \item To assess a prediction model, constructed from these characteristics.
\end{itemize}

To achieve this:
\begin{enumerate}
    \item We constructed a prediction model, where characteristics are predictor variables (X) and recall is a predicted variable (Y).
    \item We assessed the impact of individual predictor variables X using the ordinary least squares (OLS) regression model \cite{zdaniuk2024ordinary} and then assessed the prediction model using both OLS and Elastic net \cite{hans2011elastic} to cross-validate their results to assess the robustness of the model.
\end{enumerate}

For step 1, we started with construction of the prediction model. The predictor variables here are derived from the architectural, training-related, and tokenizer-related characteristics of LLMs: these characteristics and the rationale for their selection is discussed in Section~\ref{sec:background}. The values for these characteristics for each individual model are either extracted from their appropriate LLM papers or obtained empirically while interacting with these models. These characteristics and their values are presented in Table~\ref{tab:chars}. 

\begin{table*}
\renewcommand{\arraystretch}{1.5}
\centering
\caption{Characteristics of the 9 LLMs}
\label{tab:chars}
\resizebox{\textwidth}{!}{\begin{tabular}{lrrrrrrr}
\toprule
\textbf{Model} & \textbf{Training Dataset} & \textbf{Languages} & \begin{tabular}[c]{@{}r@{}}\textbf{\# Encoder} \\ \textbf{Parameters}\end{tabular} & \begin{tabular}[c]{@{}r@{}}\textbf{Embedding} \\ \textbf{Size}\end{tabular} & \begin{tabular}[c]{@{}r@{}}\textbf{Architecture} \\ \textbf{Family}\end{tabular} & \textbf{Tokenizer} \\
\midrule
CBFT & CodeSearchNet + BCB & Python, Java, JavaScript, PHP, Ruby, Go & 125 & 768 & RoBERTa & Roberta\_50265 \\
GCB & CodeSearchNet & Python, Java, JavaScript, PHP, Ruby, Go & 125 & 768 & RoBERTa & Roberta\_50265 \\
CT5 & CodeSearchNet + 2 extra languages & Python, Java, JavaScript, PHP, Ruby, Go, C, C\# & 110 & 768 & T5 & Roberta\_32100 \\
CuBERT & Github Java & Java & 345 & 1024 & BERT & FullCuBertTokenizer\_50032\\
SPTCode & CodeSearchNet & Python, Java, JavaScript, PHP, Ruby, Go & 130 & 768 & BART & Code\_50000 \\
CT5P-220 & CodeSearchNet + Github Code & Python, Java, JavaScript, PHP, Ruby, Go, C, C\#, C++ & 110 & 768 & T5 & Roberta\_32100 \\
CT5P-110 & CodeSearchNet + Github Code & Python, Java, JavaScript, PHP, Ruby, Go, C, C\#, C++ & 110 & 256 & T5 & Roberta\_32100 \\
StarEncoder & The Stack & 80+ languages, including C/C++/Java & 125 & 768 & BERT & GPT2\_49152 \\
C4 & CodeSearchNet + CodeJam/AtCoder & Python, Java, JavaScript, PHP, Ruby, Go, C\#, C++ & 125 & 768 & RoBERTa & Roberta\_50265 \\
\bottomrule
\end{tabular}
}
\end{table*}

The architecture-related characteristics are:
\begin{itemize}
    \item \textit{Architecture Family}: a nominal feature where values are the names of architectures used by the models. It is used as a composite feature that encapsulates both the type of architecture and the number of layers (see Section~\ref{sec:background}) to reduce the dimensionality of the prediction model.
    \item \textit{\# Encoder Parameters}: a numeric feature showing the number of encoder parameters in millions per a LLM.
    \item \textit{Embedding Size}: a numeric feature showing the size of a generated embedding, representing a piece of code.
\end{itemize}
The training related characteristics are:
\begin{itemize}
    \item \textit{Training Dataset}: a nominal feature where values are the names of the training datasets and their modifications. For example, for CBFT the value of this feature is 'CodeSearchNet + BCB' meaning that this model was trained on a combination of the CodeSearchNet \cite{husain2019codesearchnet} dataset and the BigCloneBench (BCB) \cite{Svajlenko2017a}. Again , this is a composite feature that encapsulates the data size and its composition to reduce dimensionality of the the prediction model.
    \item \textit{Languages}: a nominal/sequence feature where values are the programming languages used by the 'Training Dataset'.
\end{itemize}
Finally, the \textit{Tokenizer} is a nominal feature encoded as 'the name of the tokenizer'\_'the vocabulary size of the tokenizer'. For example, for CBFT LLM the tokenizer is 'Roberta\_50265' which means that Roberta tokenizer with a vocabulary size of 50265 tokens is used.

The predicted variable (Y) here is the recall, that is both numeric and continuous. The values are derived from the experiments with the 9 LLMs described in Section~\ref{subsec:rq1_design} and can be seen in the column 'Avg. Recall' in Table~\ref{tab:company_c_metrics} for Company-C dataset, in the column 'Avg. Recall' in Table~\ref{tab:company_cpp_metrics} for Company-C++ dataset, and in the column 'Avg. Recall' in Table~\ref{tbl:bcb13_rq1} for BCB13 dataset. Essentially, for every LLM in Table~\ref{tab:chars} three recall values are obtained, resulting in 27 samples, total (9 models x 3 recall values).

Following this, nominal values were transformed into a numerical format using one-shot encoding. This, however, resulted in a high feature-to-sample ratio and a very high multi-collinearity. Both led to model overfitting, poor generalization, and making regression models like OLS unsuitable. Therefore, it was decided to reduce the number of features and to simplify the existing features, as shown in Table~\ref{tbl:simplified}.

\begin{table*}[!ht]
\renewcommand{\arraystretch}{1.5}
\centering
\caption{Transformed characteristics of the 9 LLMs}
\label{tbl:simplified}
\resizebox{0.7\textwidth}{!}{
\begin{tabular}{@{}lcccccc@{}}
\toprule
\textbf{Model}       & \textbf{Training Dataset} & \textbf{Languages} & \textbf{\# Encoder Parameters} & \textbf{Embedding Size} & \textbf{Tokenizer Size} \\ \midrule
CBFT                & CodeSearchNet             & 1                  & 125                        & 768                     & 50,265                  \\
GCB                 & CodeSearchNet             & 1                  & 125                        & 768                     & 50,265                  \\
CT5                 & CodeSearchNet             & 2                  & 110                        & 768                     & 32,100                  \\
CuBERT              & Other                     & 1                  & 345                        & 1,024                   & 50,032                  \\
SPTCode             & CodeSearchNet             & 1                  & 130                        & 768                     & 50,000                  \\
CT5P-220            & CodeSearchNet             & 3                  & 110                        & 768                     & 32,100                  \\
CT5P-110            & CodeSearchNet             & 3                  & 110                        & 256                     & 32,100                  \\
StarEncoder         & Other                     & 3                  & 125                        & 768                     & 49,152                  \\
C4                  & CodeSearchNet             & 2                  & 125                        & 768                     & 50,265                  \\ \bottomrule
\end{tabular}
}
\end{table*}

In particular, the following changes have been performed:
\begin{itemize}
    \item The values in 'Training Dataset' were simplified: CodeSearchNet dataset and all its modifications were encoded as 'CodeSearchNet' and all other datasets were encoded as 'Other'. The rationale here was that modifications to CodeSearchNet are insignificant enough to avoid including them as separate entities, whereas other datasets are very different from CodeSearchNet. 
    \item The values in 'Languages' were changed so that they reflected the intersection of the languages used in the empirical studies here (C/C++/Java) and the languages in the training dataset. For example, the dataset for CBFT includes Java. Hence it shares one language with those used in the empirical studies. The rationale here was that more matching languages should have a positive impact on recall.
    \item 'Architecture Family' values and the tokenizer name values in 'Tokenizer' seemed to have a lot of overlap resulting in high collinearity between the two. Therefore, both were removed resulting in the 'Tokenizer' values becoming numerical and representing the size of the tokenizer's vocabulary. 
\end{itemize}

After applying these changes and one-shot encoding the 'Training Dataset' the feature-to-sample ratio was reduced to 0.19 (5 features / 27 samples) and also the multicollinearity of features was reduced, ranging from mild to severe for different features, yet acceptable for regression models.

In terms of investigating individual characteristics the following null hypotheses were formulated:
\begin{itemize}
    \item H01: The training dataset has no effect on the recall of LLMs.
    \item H02: The number of matching languages has no effect on the recall of LLMs.
    \item H03: The number of encoder parameters has no effect on the recall of LLMs.
    \item H04: The embedding size has no effect on the recall of LLMs.
    \item H05: The tokenizer vocabulary size has no effect on the recall of LLMs.
\end{itemize}

For step 2, we relied on OLS and Elastic net regularized regression models. The choice of regression models was due to the continuous and numeric nature of predicted variable (recall) and the smaller sample size, which makes other prediction models (for example, machine learning) less suitable. The prediction model was in acceptable format for both regression models to work: it has a continuous predicted variable (recall), smaller feature-to-sample ratio of 5 to 27, and has all nominal variables converted to numeric format. Additionally, for Elastic net, all numeric features were normalized using z-score \cite{cheadle2003analysis}. For every predictor feature $x \in X$, z-score normalization is calculated as:
\begin{equation}
z = \frac{x - \mu}{\sigma}
\label{eq:zscore_norm}
\end{equation}

where:
\begin{itemize}
    \item \(\mu\) is the mean of the feature \(x\).
    \item \(\sigma\) is the standard deviation of the feature \(x\).
    \item \(z\) is the normalized value.
\end{itemize}

The following parameter selection for Elastic net was used to address the challenges posed by a small dataset (27 samples), a high feature-to-sample ratio (5:27), and multicollinearity:
\begin{itemize}
    \item \textbf{\texttt{l1\_ratio} = [0.1, 0.5, 0.9]}: This tests different balances between L1 (Lasso) and L2 (Ridge) regularization. Lower values favor Ridge (better for dense and correlated features), while higher values increase sparsity.
    \item \textbf{\texttt{alphas} = [0.001, 0.01, 0.1, 1, 10]}: A range of regularization strengths is explored, allowing the model to select the optimal penalty level to prevent overfitting.
    \item \textbf{\texttt{cv} = 5}: Five-fold cross-validation is used to maximize the utility of limited data while ensuring model robustness.
    \item \textbf{\texttt{random\_state} = 42}: A fixed random seed ensures reproducibility of results.
\end{itemize}

This configuration ensures an optimal trade-off between bias and variance while mitigating the effects of high multicollinearity and a small sample size \cite{hans2011elastic}.

Finally, to address RQ2:
\begin{itemize}
    \item For individual features and their impact, we looked at the p-values reported by OLS for each feature and for their coefficients. This allowed to reject/accept the hypotheses.
    \item For overall model prediction, we looked at both R\^2 (the proportion of variance in the dependent variable (Y) that is explained by the independent variables (X) in the model) reported by the OLS and the Elastic net.
    \item We checked if the criteria for regression models were satisfied, such as the normality of residuals, homoscedasticity, the statistical significance of models, and the degree of multicollinearity in the independent variables (X), to ensure the robustness of the prediction model.
\end{itemize}

\subsection{Experimental design for RQ3}
To answer RQ3, we studied how effective the ensembles of the 9 LLMs trialed in this paper are, and what the effects of applying different ensembling strategies are. For this purpose, we employed the results (lists of clone candidates) originating from the RQ1 study from all four available datasets (Company-C, Company-C++, BCB13, and In-Situ-C-C++) and combined these using different ensembling strategies.

The details of empirical design common for all datasets were as follows:
\begin{enumerate}
    \item All possible combinations of 2 for the 9 given models were generated, resulting in 36 such combinations (\textbf{ensembles}), total. We didn't trial combinations of more than 2 for a couple of reasons. Firstly, increasing the size of a combination will result in a significantly higher number of combinations, making practical evaluation difficult. For example, combinations of 2 and 3 models would have resulted in 120 combinations, in total. Secondly, combinations of more than two models make analyzing their combined performance difficult, hiding individual contributions.
    \item The following \textbf{normalization methods} were adopted (see Section~\ref{sec:background}) towards the similarity score of the combined models:
    \begin{itemize}
        \item No normalization (encoded as \texttt{non-norm}): raw normalized similarity is used as generated by the LLMs.
        \item Min-Max normalization (encoded as \texttt{min-max}): similarity scores are normalized using the min-max formula, where S is similarity:
        \[
S_{\text{normalized}} = \frac{S - S_{\text{min}}}{S_{\text{max}} - S_{\text{min}}}
\]
        \item Z-score normalization (encoded as \texttt{z-score}): similarity scores are normalized as shown in Formula~\ref{eq:zscore_norm}
        \item Reciprocal Rank Fusion (encoded as \texttt{rrf}) \cite{cormack2009reciprocal} normalization: instead of using similarity scores, ranks of clone candidates are used as they appear in the sorted list of clone candidates. The formula used here is:

        \[
        \text{RRF}(r) = \frac{1}{k + r}
        \]
        
        where:
        \begin{itemize}
            \item \(r\) is the rank of the item in the list.
            \item \(k\) is a constant (here set to 60) to scale the normalization.
        \end{itemize}
    \end{itemize}
    \item When the outputs of LLMs are merged their duplicates need to be resolved (see Section~\ref{sec:background}). Three \textbf{aggregation methods} were used:
    \begin{itemize}
        \item Average duplicates' scores (encoded as \texttt{average}): here the mean similarity score of clone candidates is taken, duplicates removed retaining one copy, and the average score is assigned to the remaining clone candidate.
        \item Taking a sum of duplicates' scores (encoded as \texttt{sum}): here the sum of similarity scores of clone candidates is taken, duplicates removed retaining one copy, and the sum score is assigned to the remaining clone candidate.
        \item Taking a maximum of duplicates' scores (encoded as \texttt{max}): here the maximum of similarity scores of clone candidates is taken, duplicates removed retaining one copy, and the maximum score is assigned to the remaining clone candidate.
    \end{itemize}
    \item Combining \texttt{normalization methods} and \texttt{aggregation methods}, resulted in 12 \texttt{ensembling methods} being trialed (e.g., \texttt{non-norm\_average}.
\end{enumerate}

Equipped with this, we then proceeded in the following manner for Company-C and C++ datasets:
\begin{enumerate}
    \item All 36 combinations of LLMs were assessed for every relevant \texttt{global top K} threshold (for Company-C++ these were [10, 50, 83, 166] and for Company-C these were [10, 50, 70, 140], see Section~\ref{subsec:rq1_design}) using 12 \textbf{ensembling methods}. This resulted in 1728 candidates lists per each dataset, in total (36 combinations x 4 thresholds x 12 ensembling methods).
    \item For all 36 combinations at all \texttt{global top K} thresholds maximum individual recall \texttt{encoded as \texttt{max\_individual}} was calculated as follows: 
    \[
(M1 + M2)_{\text{max\_invividual}} = \max(Recall(M1), Recall(M2))
\]
    where M1 and M2 are two LLMs and M1 + M2 is their combination. The assumption here is that if \texttt{max\_individual} recall of two models is better than the recall of their combination then ensembling of the models does not perform better than if they were used individually. For example, if at a certain \texttt{global top K} a Model A shows a recall of 50\% and Model B shows a recall of 65\% and the recall of their combination Model A + Model B is 60\% then ensembling here does not improve recall. $$(max(50, 65) > 60)$$ Otherwise, if a recall of the combination is 70\%, for example, then ensembling here is advantageous. $$(max(50, 65) < 70)$$ 
\end{enumerate}

This approach allows us to study how effective ensembles are of all 9 LLMs on Company-C and Company-C++ datasets and also to study how effective the ensembling strategies are (ensembling methods).

Likewise, the similar approach was used for BCB13 dataset with the following adjustments:
\begin{itemize}
    \item The number of combinations was limited to the 3 best-performing (in terms of recall) combinations from Company-C and Company-C++ trials.
    \item One best-performing (again, based on Company-C and Company-C++ evaluations) ensembling method was selected.
\end{itemize}

Finally, for In-Situ-C-C++ trial, the top-performing individual models (see Section~\ref{subsec:rq1_design}) were used, resulting in 3 combinations. All 12 ensembling methods were trialled with these combinations, but here instead of recall precision was calculated and the number of clones identified (see Section~\ref{subsec:rq1_design}).

\section{Results and discussion}
\label{sec:results}
\subsection{RQ1: How effective are novel LLMs for scalable clone detection? }

\begin{table*}[h!]
\renewcommand{\arraystretch}{1.5}
\centering
\caption{Recall (\%) metrics for Company-C dataset}
\label{tab:company_c_metrics}
\resizebox{0.7\textwidth}{!}{\begin{tabular}{lccccc}
\toprule
\textbf{Model} & \textbf{Recall@10} & \textbf{Recall@50} & \textbf{Recall@70} & \textbf{Recall@140} & \textbf{Avg. Recall} \\
\midrule
CT5P-110       & 12.86              & 67.14              & 88.57              & 95.71               & 66.07               \\
SPTCode        & 14.29              & 62.86              & 80.00              & 98.57               & 63.93               \\
CuBERT         & 12.86              & 55.71              & 71.43              & 98.57               & 59.64               \\
CBFT           & 14.29              & 52.86              & 61.43              & 80.00               & 52.15               \\
CT5            & 12.86              & 50.00              & 64.29              & 81.43               & 52.15               \\
GCB            & 12.86              & 51.43              & 60.00              & 77.14               & 50.36               \\
StarEncoder    & 12.86              & 42.86              & 60.00              & 77.14               & 48.22               \\
CT5P-220       & 12.86              & 38.57              & 40.00              & 51.43               & 35.72               \\
C4             & 14.29              & 38.57              & 38.57              & 48.57               & 35.00               \\
\bottomrule
\end{tabular}}
\end{table*}

\begin{table*}[h!]
\renewcommand{\arraystretch}{1.5}
\centering
\caption{Recall (\%) metrics for Company-C++ dataset}
\label{tab:company_cpp_metrics}
\resizebox{0.7\textwidth}{!}{\begin{tabular}{lccccc}
\toprule
\textbf{Model} & \textbf{Recall@10} & \textbf{Recall@50} & \textbf{Recall@83} & \textbf{Recall@166} & \textbf{Avg. Recall} \\
\midrule
CT5P-110       & 10.84              & 55.42              & 83.13              & 98.80               & 62.05               \\
CuBERT         & 10.84              & 51.81              & 78.31              & 96.39               & 59.34               \\
SPTCode        & 10.84              & 51.81              & 77.11              & 92.77               & 58.13               \\
CT5            & 8.43               & 46.99              & 78.31              & 92.77               & 56.63               \\
StarEncoder    & 9.64               & 50.60              & 74.70              & 90.36               & 56.33               \\
GCB            & 9.64               & 50.60              & 75.90              & 84.34               & 55.12               \\
CBFT           & 10.84              & 53.01              & 71.08              & 81.93               & 54.22               \\
CT5P-220       & 7.23               & 44.58              & 66.27              & 77.11               & 48.80               \\
C4             & 9.64               & 33.73              & 42.17              & 50.60               & 34.04               \\
\bottomrule
\end{tabular}}
\end{table*}

\begin{table*}[!th]
\renewcommand{\arraystretch}{1.5}
\caption{Recall (\%) metrics for BCB13 dataset}
\label{tbl:bcb13_rq1}
\centering
\resizebox{0.6\textwidth}{!}{\begin{tabular}{lccccccc}
\toprule
\textbf{Model}      & \textbf{T1} & \textbf{T2} & \textbf{VST3} & \textbf{ST3} & \textbf{MT3} & \textbf{WT3/T4} & \textbf{Avg. Recall} \\ 
\midrule
CT5                 & 100         & 97          & 98           & 90           & 39           & 1              & 70.83                \\ 
StarEncoder         & 100         & 97          & 98           & 89           & 38           & 1              & 70.50                \\ 
CuBERT              & 100         & 97          & 93           & 86           & 35           & 1              & 68.66                \\ 
SPTCode             & 100         & 97          & 97           & 83           & 28           & 1              & 67.66                \\ 
GCB                 & 100         & 97          & 96           & 80           & 27           & 1              & 66.83                \\ 
CBFT                & 100         & 96          & 96           & 77           & 26           & 1              & 66.00                \\ 
CT5P-110            & 100         & 94          & 93           & 75           & 20           & 1              & 63.83                \\ 
CT5P-220            & 100         & 94          & 84           & 67           & 24           & 1              & 61.67                \\ 
C4                  & 100         & 92          & 79           & 45           & 10           & 1              & 54.50                \\ 
\bottomrule
\end{tabular}}
\end{table*}

\begin{table}[h!]
\renewcommand{\arraystretch}{1.5}
\centering
\caption{Selecting top-performing models: ranking of models using Borda's count}
\label{tab:ranking_models}
\resizebox{0.5\textwidth}{!}{\begin{tabular}{lccccl}
\toprule
\multirow{2}{*}{\textbf{Model}} & \multicolumn{3}{c}{\textbf{Ranking (Borda count)}} & \multirow{2}{*}{\begin{tabular}[c]{@{}r@{}}\textbf{Total} \\ \textbf{Borda count}\end{tabular}} & \multirow{2}{*}{\begin{tabular}[c]{@{}r@{}}\textbf{St.Dev of} \\ \textbf{Ranking}\end{tabular}} \\ \cmidrule{2-4}
                                & \textbf{C}       & \textbf{C++}     & \textbf{BCB13}   &                                  &                                  \\ \midrule
CuBERT         & 3 (7)                & 2 (8)                & 3 (7)                & 22                         & 0.58                       \\
CT5P-110       & 1 (9)                & 1 (9)                & 7 (3)                & 21                         & 3.46                       \\
SPTCode        & 2 (8)                & 3 (7)                & 4 (6)                & 21                         & 1.00                       \\
CT5            & 4 (6)                & 4 (6)                & 1 (9)                & 21                         & 1.73                       \\
StarEncoder    & 6 (4)                & 5 (5)                & 2 (8)                & 17                         & 2.08                       \\
GCB            & 5 (5)                & 6 (4)                & 5 (5)                & 14                         & 0.58                       \\
CBFT           & 4 (6)                & 7 (3)                & 6 (4)                & 13                         & 1.53                       \\
CT5P-220       & 7 (3)                & 8 (2)                & 8 (2)                & 7                          & 0.58                       \\
C4             & 8 (2)                & 9 (1)                & 9 (1)                & 4                          & 0.58                       \\ \bottomrule
\end{tabular}

}
\end{table}

\begin{table}[h!]
\renewcommand{\arraystretch}{1.5}
\centering
\caption{Symmetric Differences Between Model Pairs}
\label{tab:symmetric_diff}
\resizebox{0.3\textwidth}{!}{\begin{tabular}{llr}
\toprule
\textbf{Model 1}   & \textbf{Model 2}   & \textbf{Symmetric Diff} \\ 
\midrule
SPTCode            & CuBERT             & 4,559,188               \\
CT5P-110               & CuBERT             & 4,717,994               \\
CT5P-110               & SPTCode            & 4,809,477               \\
StarEncoder        & CT5P-110               & 4,808,409               \\
SPTCode            & StarEncoder        & 4,544,510               \\
CuBERT             & StarEncoder        & 4,342,901               \\
\bottomrule
\end{tabular}}
\end{table}

\begin{table}[h!]
\renewcommand{\arraystretch}{1.5}
\centering
\caption{Precision of models in an in-situ evaluation}
\label{tab:true_clones_precision}
\resizebox{0.3\textwidth}{!}{\begin{tabular}{lrr}
\toprule
\textbf{Model}        & \textbf{\# True Clones} & \textbf{Precision (\%)} \\ 
\midrule
CBFT            & 213                     & 15.32                    \\
CuBERT                & 251                     & 18.06                    \\
CT5P-110              & 552                     & 39.71                    \\ 
\bottomrule
\end{tabular}}
\end{table}

To address RQ1, we evaluated nine novel LLMs across three public datasets: Company-C, Company-C++, and BCB13 using a scalable, function-level batch search strategy, and subsequently selected the top performers for in-situ evaluation on the company’s private In-Situ-C-C++ dataset.

\subsubsection{Results on Public Datasets}

Table~\ref{tab:company_c_metrics} summarizes the recall metrics for the Company-C dataset arranged by the ``Avg. Recall'' values. Notably, CT5P-110 achieved the highest average recall of 66.07\%, with recalls of 12.86, 67.14, 88.57, and 95.71 at \texttt{global top K} values of 10, 50, 70, and 140, respectively. SPTCode and CuBERT were second and third and also yielded competitive performance, with average recalls of 63.93\% and 59.64\%, respectively. For the Company-C++ dataset (see Table~\ref{tab:company_cpp_metrics}), a similar trend was observed. CT5P-110 again led with an average recall of 62.05\%, followed by CuBERT (59.34\%) and SPTCode (58.13\%). The two worst performing models for both these datasets were CT5P-220 and C4.

In contrast, evaluation on the BCB13 dataset (see Table~\ref{tbl:bcb13_rq1}) revealed a shift in performance patterns. Here, models such as CT5 and StarEncoder achieved the highest average recalls of 70.83\% and 70.50\%, respectively, while CT5P-110’s average recall was lower at 63.83\% (ranking 7th). Interestingly though, the worst performing models here were CT5P-220 and C4, similar to Company-C and Company-C++ trials.

These results suggest that CT5P-110 is highly effective in scenarios where the codebase is of moderate size and clone class distributions are relatively constrained. But the distinctive nature of BCB13—characterized by large clone classes and Java probably accounts, to a large degree, for its lesser ranking and the altered rankings of other models.

\subsubsection{Aggregated Ranking Analysis}

To aggregate performance across the three datasets, we employed the Borda count method, with the rankings summarized in Table~\ref{tab:ranking_models}. According to the data presented in this table several trends could be highlighted:
\begin{itemize}
    \item Low St.Dev ($\leq$ 1.0) models (CuBERT, GCB, CT5P-220, C4, SPTCode) demonstrate reliable and consistent performance. These models maintain similar rankings across datasets. CuBERT is a strong, stable performer (always top 3) whereas C4 and CT5P-220 are consistently poor performers.
    \item Moderate St.Dev (1.5 - 2.5) models (CT5, CBFT, StarEncoder) appear to be situationally strong and seem to be dataset-sensitive.
    \item High St.Dev ($\geq$ 3.0) model CT5P-110 shows highly inconsistent performance: it performs exceptionally well (1st place) in Company-C and Company-C++ datasets, but poorly (7th place) in BCB13. This might suggest specialization or overfitting to specific dataset characteristics.
\end{itemize}

If a reliable all-round model is needed CuBERT looks like the most consistent model, obtaining the highest total Borda count (22), with a low standard deviation (0.58), indicating stable performance across Company-C, Company-C++, and BCB13 datasets. CT5P-110 can be a high-risk, high-reward model: although CT5P-110 ranked first on both Company-C and Company-C++ datasets, its lower ranking (7th) on BCB13 resulted in a total Borda count of 21 and a higher variability (standard deviation of 3.46). CT5 and StarEncoder (good rankings but varied performance) seem to excel in certain datasets. GCB, C4, and CT5P-220 are consistently ranked low but are predictable and could be considered if stability is a priority, even if performance is low.

\subsubsection{In-Situ Evaluation}

The final phase involved selecting the top-performing LLMs for in-situ evaluation on the in-situ-C-C++ dataset using the criteria outlined in Section~\ref{subsec:rq1_design}. In particular, we have selected:
\begin{itemize}
    \item The CuBERT model: this is the top-performing and the most predictable model with a Borda count of 22 and low St.Dev of rankings (0.58), as can be seen in Table~\ref{tab:ranking_models}.
    \item The CT5P-110 model: this is the second top ranking model (along with CT5 and SPTCode) achieving a Borda count of 21 (see Table~\ref{tab:ranking_models}). This model can produce more unique clone candidates, as evidenced by the symmetric difference in Table~\ref{tab:symmetric_diff}: The table shows that, in all combinations where this model was present, including CuBERT (which was also brought forward for this evaluation), the highest symmetric differences were achieved, suggesting higher uniqueness for ensembling.
\end{itemize}

Table~\ref{tab:true_clones_precision} shows the results of an in-situ evaluation with the in-Situ-C-C++ dataset. Here CT5P-110 detected 552 true clones with a precision of 39.71\%, markedly outperforming both CBFT (213 true clones, 15.32\% precision) and CuBERT (251 true clones, 18.06\% precision). Overall, the CT5P model outperforms CBFT by 159.2\% and the CuBERT model by 119\% in terms of precision. Because of the (top-1390-candidate-pairs) manner in which the data was gathered, the improvement in the number of clones identified is the same. Despite its mixed performance across public datasets, CT5P-110’s superior precision in a realistic operational setting on C/C++ code underscores its practical effectiveness for scalable clone detection. 

\subsubsection{Discussion}

These experimental outcomes yield several key insights:

\begin{itemize} 
\item \textbf{Dataset Dependency:} The performance of the LLMs is markedly dataset-dependent. While CT5P-110 excelled on the Company-C and Company-C++ datasets, its performance on BCB13 was comparatively lower. This suggests that factors such as language and/or clone-class size influence model effectiveness; the Company-C, Company-C++, and In-Situ-C-C++ datasets all have smaller clone classes (see Section~\ref{subsec:rq1_design}).
\item \textbf{Stability Versus Peak Performance:} The Borda count aggregation (see Table~\ref{tab:ranking_models}) highlights the importance of consistency. Although CT5P-110 achieved top rankings on two datasets, its lower performance on BCB13 increased variability. In contrast, CuBERT’s stable performance across datasets makes it a compelling candidate for scenarios where uniformity is critical.
\item \textbf{Real-World Applicability:} The in-situ evaluation (see Table~\ref{tab:true_clones_precision}) is particularly illuminating. CT5P-110’s ability to detect a substantially higher number of true clones with significantly better precision in the private dataset demonstrates that, public benchmark performance may not always be an absolute guide for private-code performance.
\end{itemize}

In summary with respect to RQ1, this evaluation reveals that, while no single LLM uniformly dominates across all datasets, CT5P-110 and CuBERT emerge as strong candidates, followed closely by CT5 and SPTCode. CT5P-110, in particular, shows significant promise in real-world, in-situ applications despite some variability in public benchmarks.

\subsection{RQ2: How do characteristics of these LLMs affect their effectiveness with regard to recall?}

\subsubsection{Regression Analysis Results}
To investigate the influence of model characteristics on clone detection recall, we constructed a prediction model using both Ordinary Least Squares (OLS) and Elastic Net regression. The predictor variables comprised architectural (number of encoder parameters, embedding size), training-related (training dataset, quantified as a binary indicator with \texttt{Other} vs. CodeSearchNet, and number of matching languages), and tokenizer-related characteristics (tokenizer vocabulary size) (see Table~\ref{tbl:simplified}).

For the Elastic Net model, hyperparameter tuning via five-fold cross-validation resulted in an optimal configuration of \(\alpha=0.1\) and \(\texttt{l1\_ratio}=0.1\). The corresponding coefficients were:
\begin{itemize}
    \item \textbf{Languages}: \(-6.27\)
    \item \textbf{\# Encoder Parameters}: \(-0.49\)
    \item \textbf{Embedding Size}: \(-4.42\)
    \item \textbf{Tokenizer Vocabulary Size}: \(-3.55\)
    \item \textbf{Training Dataset (Other)}: \(+5.33\)
\end{itemize}
This model explained approximately 34.7\% of the variance in recall (\(R^2 = 0.347\)).

The OLS regression analysis yielded an \(R^2\) of 0.445 (adjusted \(R^2 = 0.313\)) with the following estimated coefficients and p-values:
\begin{itemize}
    \item \textbf{Languages} (\(x_1\)): Coefficient = \(-13.40\), \(p = 0.002\)
    \item \textbf{\# Encoder Parameters} (\(x_2\)): Coefficient = \(-5.39\), \(p = 0.134\)
    \item \textbf{Embedding Size} (\(x_3\)): Coefficient = \(-6.41\), \(p = 0.013\)
    \item \textbf{Tokenizer Vocabulary Size} (\(x_4\)): Coefficient = \(-8.12\), \(p = 0.014\)
    \item \textbf{Training Dataset (Other)} (\(x_5\)): Coefficient = \(+12.11\), \(p = 0.005\)
\end{itemize}

\subsubsection{Model Diagnostics and Robustness}
To ensure the validity of the regression models, we performed standard diagnostic checks, including normality of residuals, homoscedasticity, and multicollinearity assessment.

\textbf{Normality of Residuals:} In OLS, the Jarque-Bera test resulted in a test statistic of 1.751 with a -value of 0.417, indicating that the residuals do not significantly deviate from normality. Additionally, the skewness of the residuals was 0.307, and the kurtosis was 1.914, both within an acceptable range for linear regression.

\textbf{Homoscedasticity:} The Breusch-Pagan test yielded a test statistic of 10.21 with a -value of 0.037, indicating the presence of heteroscedasticity. This suggests that variance of residuals is not constant, and robust standard errors (HC3) were used in OLS to mitigate this issue.

\textbf{Multicollinearity:} The OLS model's condition number was 6.56, indicating that multicollinearity is not a concern and that the regression coefficients are stable.

\textbf{Durbin-Watson Test:} The Durbin-Watson statistic was 1.004, suggesting some positive autocorrelation in the residuals, but given the small dataset size (27 observations), this effect is not critical.

\textbf{Statistical Significance of the Model:} The overall F-statistic in OLS was 3.370 with a -value of 0.0217, indicating that at least one predictor variable significantly contributes to explaining the variance in recall.

These diagnostics confirm that the OLS regression model satisfies key robustness assumptions and provides robust, interpretable insights into the relationships between model characteristics and recall performance.

\subsubsection{Hypothesis Testing}
Based on the OLS p-values, we evaluate the following null hypotheses:
\begin{itemize}
    \item \textbf{H01:} The training dataset has no effect on the recall of LLMs. \\
    \textbf{Decision:} Rejected (\(p=0.005\)). The significant positive coefficient for the \texttt{Training Dataset (Other)} variable indicates that models trained on datasets other than CodeSearchNet yield higher recall.
    
    \item \textbf{H02:} The number of matching languages has no effect on the recall of LLMs. \\
    \textbf{Decision:} Rejected (\(p=0.002\)). The significant negative coefficient suggests that an increase in the number of matching languages is associated with lower recall, contrary to the initial expectation.
    
    \item \textbf{H03:} The number of encoder parameters has no effect on the recall of LLMs. \\
    \textbf{Decision:} Accepted (\(p=0.134\)). The effect of encoder parameters is not statistically significant.
    
    \item \textbf{H04:} The embedding size has no effect on the recall of LLMs. \\
    \textbf{Decision:} Rejected (\(p=0.013\)). A larger embedding size is significantly associated with lower recall.
    
    \item \textbf{H05:} The tokenizer vocabulary size has no effect on the recall of LLMs. \\
    \textbf{Decision:} Rejected (\(p=0.014\)). The tokenizer vocabulary size also shows a significant negative effect.
\end{itemize}

\subsubsection{Discussion}

The regression results indicate that most of the investigated characteristics significantly influence clone detection recall. The negative coefficients for \textit{Languages}, \textit{Embedding Size}, and \textit{Tokenizer Vocabulary Size} imply that increases in these features are associated with a decrease in recall. The negative effect of more matching languages contradicts the initial expectation that more matching languages would improve recall. A possible explanation here could be that more matching languages in general might increase tokenization inconsistencies or model overfitting. Larger embedding sizes negatively affect recall possibly because too large embeddings dilute useful information for clone detection. Likewise, a larger tokenizer vocabulary might lead to more fragmented tokenization, reducing the model's ability to recognize clones, but these are important finding, based on their counter-intuitive nature and deserve further buttressing by other researchers. Conversely, the positive coefficient for the \textit{Training Dataset (Other)} variable indicates that models trained on datasets different from CodeSearchNet tend to perform better in clone detection tasks. This could reflect differences in the data composition or training strategies that are more effective to generalizing on clone detection.

The non-significant effect of the number of encoder parameters suggests that simply scaling model size in terms of parameters does not necessarily improve recall performance. This finding, along with the Training-Dataset finding, is in-line with the recent trends in LLM research focusing on improving training datasets and process rather than the size of the LLMs \cite{hoffmann2022training}.

Overall, an F-statistic of 3.370 with p-value of 0.0217 indicates that the OLS model, as a whole, is statistically significant, meaning at least one predictor has a meaningful impact on recall. The OLS model (\(R^2\) of 0.445 (adjusted \(R^2 = 0.313\)) explains more variance in recall but may overfit due to lack of regularization, as indicated by the drop in adjusted \(R^2\). In contrast, Elastic Net (\(R^2\) of 0.3472) sacrifices some fit but is likely more generalizable. Both models explain a moderate amount of variance (35-45\%), suggesting that other important factors influence recall. The relatively low adjusted \(R^2\) implies that some predictors may contribute noise rather than useful signal. However, given the small dataset (27 samples), Elastic Net is likely the more reliable model since it prevents overfitting, while OLS provides better interpretability. Further improvements, such as adding relevant features or interaction terms, could enhance this predictive power.

Although Elastic Net provides better reliability by mitigating overfitting, the OLS model was chosen for detailed presentation primarily due to its greater interpretability, allowing clearer insights into predictor contributions, which is equally important for our specific research context. We acknowledge this choice explicitly: while Elastic Net may be more robust, interpretability was prioritized here to gain transparent insights from predictors. Future work could explicitly include Elastic Net results or adopt regularization to balance interpretability and generalizability more effectively.

Answering RQ2, the effectiveness of LLMs in clone detection (measured by recall) seems to be strongly influenced by training data selection, tokenizer configuration, and embedding size. Larger models (in terms of parameters) do not necessarily improve recall, challenging the assumption that increasing model size alone enhances performance and in line with recent LLM development trends \cite{hoffmann2022training}. These findings suggest that choosing the right dataset and optimizing the tokenization process are more critical than simply increasing model complexity. Moderate \(R^2\) values indicate that there are likely other important characteristics at play.

\subsection{RQ3: How effective are the ensembles of these LLMs? }

% \begin{table}[ht]
% \renewcommand{\arraystretch}{1.5}
% \centering
% \caption{Average recall for LLM ensembles in Company-C dataset}
% \label{tab:ensembles-combinations-c}
% \resizebox{\columnwidth}{!}{\input{tables/ensemble-combinations-c}}
% \end{table}

\begin{table}[ht]
\renewcommand{\arraystretch}{1.5}
\centering
\caption{Average recall for LLM ensembles in Company-C dataset}
\label{tab:ensembles-combinations-c}
\resizebox{0.5\textwidth}{!}{\begin{tabular}{lrrrl}
\toprule
   Ensemble &  Avg. Ind & Avg. Recall of & Best  & Best Ensembling \\
            & Max Recall & ensembling & Ensembling & Method\\
            && methods & Recall & \\
\midrule
CT5P-110 + SPTCode & 67.14 & 66.55 & \textit{67.14} & non-norm\_sum \\
CT5P-110 + CuBERT & 66.79 & 64.58 & 66.07 & non-norm\_sum \\
CT5P-110 + CT5 & 66.07 & 61.90 & \textit{66.07} & z-score\_max \\
CT5P-110 + CBFT & 66.43 & 61.49 & 66.07 & z-score\_max \\
SPTCode + CuBERT & 63.93 & 61.37 & \textbf{64.29} & z-score\_max \\
CT5P-110 + GCB & 66.07 & 60.83 & \textit{66.07} & z-score\_max \\
SPTCode + CT5 & 63.93 & 60.21 & 63.21 & z-score\_max \\
CT5P-110 + StarEncoder & 66.07 & 60.00 & \textit{66.07} & non-norm\_sum \\
SPTCode + CBFT & 63.93 & 59.14 & \textit{63.93} & z-score\_max \\
SPTCode + GCB & 63.93 & 57.62 & 63.57 & z-score\_max \\
CT5P-110 + CT5P-220 & 66.07 & 56.82 & \textit{66.07} & z-score\_max \\
CuBERT + CBFT & 60.00 & 56.25 & 59.64 & z-score\_max \\
CuBERT + CT5 & 59.64 & 56.16 & \textit{59.64} & non-norm\_max \\
CT5P-110 + C4 & 66.43 & 55.92 & 65.71 & non-norm\_max \\
StarEncoder + CuBERT & 59.64 & 55.57 & 58.93 & non-norm\_sum \\
GCB + CuBERT & 59.64 & 55.48 & 58.93 & z-score\_max \\
StarEncoder + SPTCode & 63.93 & 55.45 & 63.57 & z-score\_max \\
C4 + SPTCode & 63.93 & 54.26 & 61.79 & non-norm\_max \\
SPTCode + CT5P-220 & 63.93 & 53.39 & 63.57 & z-score\_max \\
GCB + CBFT & 52.14 & \textbf{52.62} & \textbf{53.21} & non-norm\_max \\
StarEncoder + GCB & 50.36 & \textbf{51.70} & \textbf{53.93} & min-max\_max \\
CT5 + CBFT & 53.21 & 51.70 & 52.50 & z-score\_max \\
StarEncoder + CBFT & 52.14 & 51.61 & \textbf{53.21} & rrf\_max \\
GCB + CT5 & 52.50 & 51.10 & \textbf{53.21} & rrf\_max \\
CuBERT + CT5P-220 & 59.64 & 50.60 & \textbf{60.00} & non-norm\_max \\
StarEncoder + CT5 & 52.14 & 49.58 & \textbf{53.21} & z-score\_max \\
C4 + CuBERT & 60.00 & 47.71 & \textbf{60.71} & non-norm\_max \\
CT5P-220 + CT5 & 52.14 & 46.93 & 51.07 & z-score\_max \\
CT5P-220 + CBFT & 52.14 & 46.01 & \textbf{52.86} & min-max\_max \\
C4 + CT5 & 52.50 & 45.33 & 51.43 & non-norm\_max \\
GCB + CT5P-220 & 50.36 & 45.30 & \textit{50.36} & non-norm\_max \\
C4 + CBFT & 52.14 & 43.84 & \textit{52.14} & non-norm\_max \\
StarEncoder + CT5P-220 & 48.21 & 43.39 & \textbf{49.64} & min-max\_max \\
C4 + GCB & 50.71 & 42.86 & \textbf{51.07} & non-norm\_max \\
C4 + StarEncoder & 48.57 & 42.77 & \textbf{50.00} & non-norm\_max \\
C4 + CT5P-220 & 36.07 & \textbf{39.35} & \textbf{44.64} & rrf\_sum \\
\bottomrule
\end{tabular}
}
\end{table}

% \begin{table}[ht]
% \renewcommand{\arraystretch}{1.5}
% \centering
% \caption{Average recall for LLM ensembles in Company-C++ dataset}
% \label{tab:ensembles-combinations-cpp}
% \resizebox{\columnwidth}{!}{\input{tables/ensemble-combinations-cpp}}
% \end{table}

\begin{table}[ht]
\renewcommand{\arraystretch}{1.5}
\centering
\caption{Average recall for LLM ensembles in Company-C++ dataset}
\label{tab:ensembles-combinations-cpp}
\resizebox{0.5\textwidth}{!}{\begin{tabular}{lrrrl}
\toprule
   Ensemble &  Avg. Ind & Avg. Recall of & Best  & Best Ensembling \\
            & Max Recall & ensembling & Ensembling & Method\\
            && methods & Recall & \\
\midrule
CT5P-110 + SPTCode & 62.05 & 61.67 & \textbf{62.35} & non-norm\_max \\
CT5P-110 + CuBERT & 62.05 & 61.04 & 62.05 & non-norm\_sum \\
CT5P-110 + CT5 & 62.05 & 60.07 & 62.05 & z-score\_max \\
SPTCode + CuBERT & 59.34 & \textbf{59.69} & \textbf{61.75} & rrf\_sum \\
GCB + CuBERT & 59.34 & \textbf{59.66} & \textbf{60.24} & non-norm\_sum \\
CuBERT + CBFT & 59.64 & 59.14 & \textbf{60.24} & non-norm\_max \\
StarEncoder + CuBERT & 59.34 & 59.11 & \textbf{60.54} & min-max\_max \\
CT5P-110 + GCB & 62.05 & 58.73 & 62.05 & z-score\_max \\
CuBERT + CT5 & 59.34 & 58.58 & \textbf{59.94} & rrf\_max \\
CT5P-110 + CBFT & 62.05 & 58.43 & 62.05 & z-score\_max \\
SPTCode + CT5 & 58.43 & 57.96 & 58.43 & non-norm\_sum \\
CT5P-110 + StarEncoder & 62.05 & 57.13 & 62.05 & non-norm\_sum \\
StarEncoder + GCB & 56.63 & \textbf{57.08} & \textbf{58.73} & rrf\_sum \\
CuBERT + CT5P-220 & 59.34 & 56.68 & 59.34 & z-score\_max \\
CT5P-110 + CT5P-220 & 62.05 & 56.53 & 62.05 & z-score\_max \\
StarEncoder + CBFT & 57.23 & 56.45 & \textbf{59.04} & rrf\_max \\
GCB + CT5 & 57.83 & 56.10 & 57.83 & non-norm\_sum \\
CT5 + CBFT & 58.73 & 55.97 & 57.23 & non-norm\_sum \\
GCB + CBFT & 56.02 & 55.65 & \textbf{56.33} & min-max\_sum \\
SPTCode + CBFT & 58.43 & 55.42 & \textbf{59.34} & rrf\_sum \\
StarEncoder + CT5 & 57.83 & 55.17 & 57.53 & min-max\_max \\
SPTCode + GCB & 58.13 & 54.99 & \textbf{58.73} & non-norm\_sum \\
SPTCode + CT5P-220 & 58.13 & 54.47 & 58.13 & z-score\_max \\
CT5P-220 + CT5 & 56.63 & 53.89 & 56.63 & z-score\_max \\
StarEncoder + SPTCode & 58.13 & 53.39 & \textbf{58.43} & non-norm\_sum \\
CT5P-220 + CBFT & 54.22 & 52.94 & \textbf{55.72} & rrf\_sum \\
GCB + CT5P-220 & 55.12 & 52.91 & \textbf{55.72} & rrf\_sum \\
StarEncoder + CT5P-220 & 56.33 & 52.26 & \textbf{56.63} & z-score\_max \\
CT5P-110 + C4 & 62.05 & 51.48 & 62.05 & z-score\_max \\
C4 + CuBERT & 59.34 & 50.95 & \textbf{59.94} & non-norm\_max \\
C4 + CT5 & 56.93 & 49.20 & 56.33 & z-score\_max \\
C4 + SPTCode & 58.13 & 48.37 & 58.13 & z-score\_max \\
C4 + GCB & 55.12 & 47.72 & 55.12 & non-norm\_max \\
C4 + CBFT & 54.22 & 46.64 & 53.61 & non-norm\_max \\
C4 + StarEncoder & 56.33 & 44.60 & \textbf{57.23} & min-max\_max \\
C4 + CT5P-220 & 49.40 & 41.87 & 49.40 & non-norm\_max \\
\bottomrule
\end{tabular}

}
\end{table}

\begin{table}[ht]
\renewcommand{\arraystretch}{1.5}
\centering
\caption{Average recall for ensembling methods in Company-C dataset}
\label{tab:ensembles-c}
\resizebox{0.3\textwidth}{!}{\begin{tabular}{lr}
\toprule
\textbf{Ensembling method} & \textbf{Avg. Recall (\%)} \\
\midrule
max\_individual        & 58.63 \\ \hline
min-max\_max           & 56.77 \\
rrf\_max               & 55.55 \\
z-score\_max           & 55.05 \\
rrf\_sum               & 54.75 \\
non-norm\_max          & 54.56 \\
z-score\_sum           & 53.89 \\
non-norm\_sum          & 53.61 \\
min-max\_sum           & 53.57 \\
rrf\_average           & 52.92 \\
min-max\_average       & 52.69 \\
z-score\_average       & 51.63 \\
non-norm\_average      & 45.98 \\
\bottomrule
\end{tabular}
}
\end{table}

\begin{table}[ht]
\renewcommand{\arraystretch}{1.5}
\centering
\caption{Average recall for ensembling methods in Company-C++ dataset}
\label{tab:ensembles-cpp}
\resizebox{0.3\textwidth}{!}{\begin{tabular}{lr}
\toprule
\textbf{Ensembling method} & \textbf{Avg. Recall (\%)} \\
\midrule
max\_individual      & 58.83 \\ \hline
rrf\_sum            & 57.26 \\
z-score\_max        & 57.20 \\
z-score\_sum        & 56.80 \\
non-norm\_max       & 56.33 \\
rrf\_max            & 56.31 \\
min-max\_max        & 56.20 \\
min-max\_sum        & 55.47 \\
non-norm\_sum       & 55.08 \\
rrf\_average        & 54.90 \\
min-max\_average    & 52.10 \\
z-score\_average    & 50.47 \\
non-norm\_average   & 47.68 \\
\bottomrule
\end{tabular}
}
\end{table}

\begin{figure*}[t]
    \centering
    % First Heatmap
    \begin{minipage}{0.48\textwidth}
        \centering
        \includegraphics[width=\linewidth]{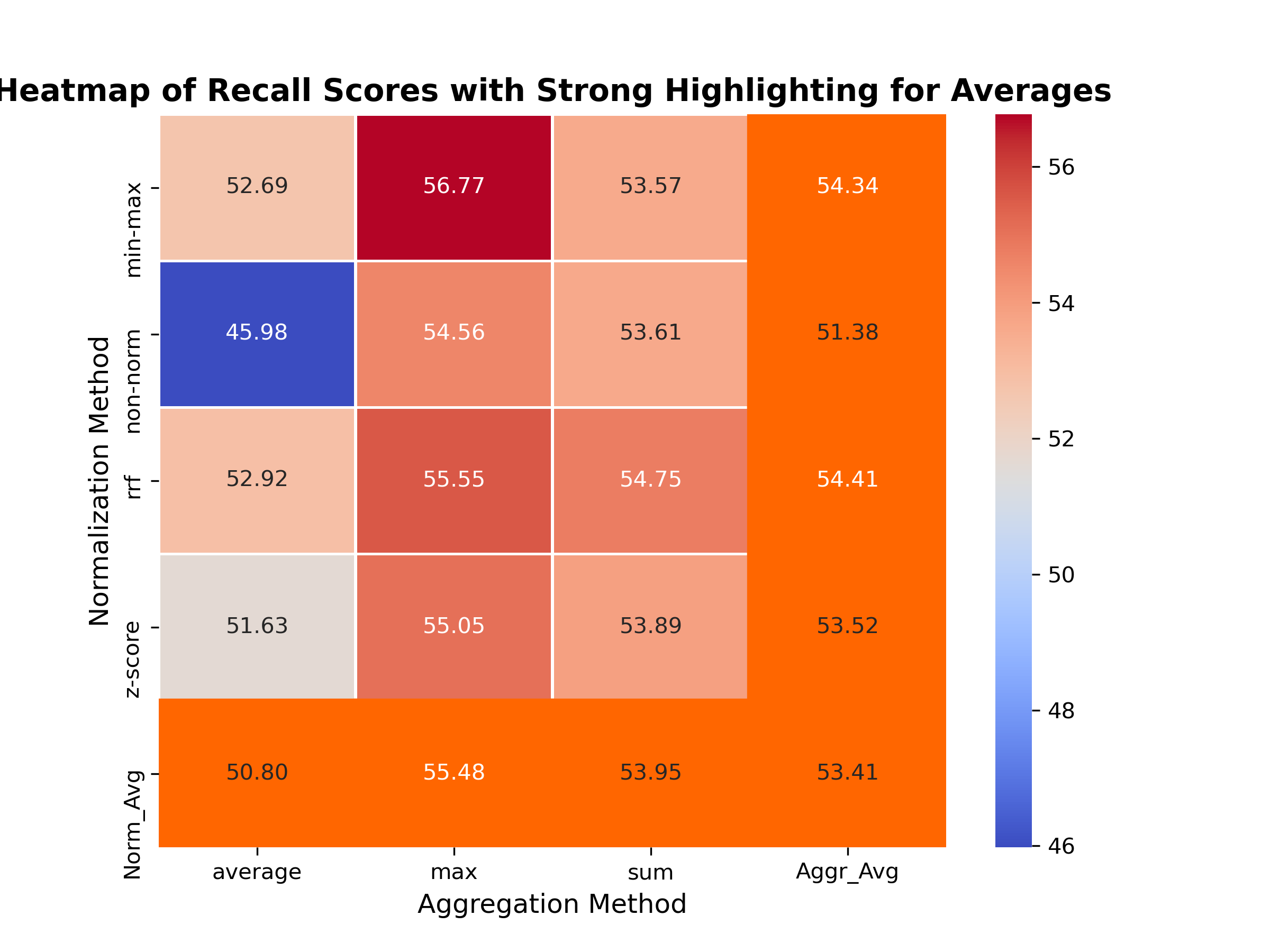}
        \caption{Avg. Recall by Normalization \& Aggregation for Company-C}
        \label{fig:heatmap1}
    \end{minipage}
    \hfill
    % Second Heatmap
    \begin{minipage}{0.48\textwidth}
        \centering
        \includegraphics[width=\linewidth]{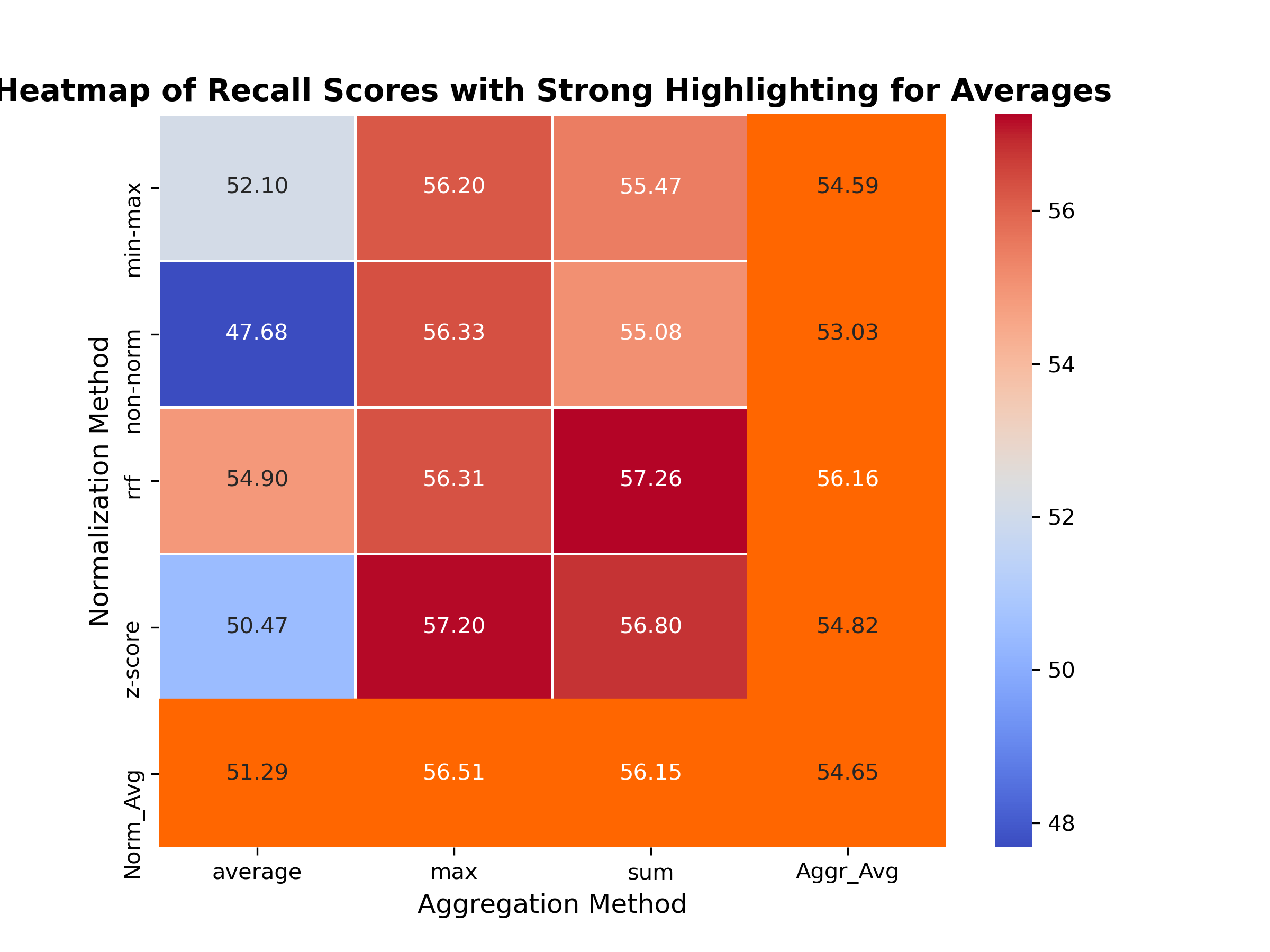}
        \caption{Avg. Recall by Normalization \& Aggregation for Company-C++}
        \label{fig:heatmap2}
    \end{minipage}
\end{figure*}

% \begin{figure*}[t]
%     \centering
%     % First Heatmap
%     \includegraphics[width=0.8\linewidth]{figures/heatmap-c.png}
%     \caption{Avg. Recall by Normalization \& Aggregation for Company-C}
%     \label{fig:heatmap1}

%     \vspace{1em} % Add some vertical spacing

%     % Second Heatmap
%     \includegraphics[width=0.8\linewidth]{figures/heatmap-cpp.png}
%     \caption{Avg. Recall by Normalization \& Aggregation for Company-C++}
%     \label{fig:heatmap2}
% \end{figure*}

\begin{table*}[h]
\renewcommand{\arraystretch}{1.5}
\centering
\caption{Comparison of recall for select ensembles for BCB13 dataset}
\label{tab:ensemble_results_bcb}
\resizebox{\textwidth}{!}{
\begin{tabular}{l l r r r r r r r r r}
\toprule
\textbf{Ensemble} & \textbf{Ensembling Method} & \textbf{Type-1} & \textbf{Type-2} & \textbf{VST3} & \textbf{ST3} & \textbf{MT3} & \textbf{WT3/T4} & \textbf{Avg Recall} & \textbf{Ind Max} & \textbf{Diff} \\
\midrule
CT5P-110 + CT5 & min-max (max) & 100 & 99 & 98 & 91 & 40 & 1 & 71.50 & 70.83 & 0.67 \\
CT5P-110 + CT5 & min-max (sum) & 100 & 99 & 98 & 91 & 40 & 1 & 71.50 & 70.83 & 0.67 \\
CT5P-110 + CT5 & z-score (max) & 100 & 99 & 98 & 91 & 40 & 1 & 71.50 & 70.83 & 0.67 \\
CT5P-110 + CT5 & rrf (max) & 100 & 99 & 98 & 91 & 40 & 1 & 71.50 & 70.83 & 0.67 \\
CT5P-110 + CT5 & rrf (sum) & 100 & 99 & 98 & 91 & 40 & 1 & 71.50 & 70.83 & 0.67 \\
CT5P-110 + CT5 & z-score (sum) & 100 & 99 & 98 & 88 & 38 & 1 & 70.67 & 70.83 & \textit{-0.16} \\
CT5P-110 + CuBERT & min-max (max) & 100 & 99 & 97 & 88 & 36 & 2 & 70.33 & 68.66 & 1.67 \\
CT5P-110 + CuBERT & min-max (sum) & 100 & 99 & 97 & 88 & 36 & 2 & 70.33 & 68.66 & 1.67 \\
CT5P-110 + CuBERT & z-score (max) & 100 & 99 & 97 & 88 & 36 & 2 & 70.33 & 68.66 & 1.67 \\
CT5P-110 + CuBERT & rrf (max) & 100 & 99 & 97 & 88 & 36 & 2 & 70.33 & 68.66 & 1.67 \\
CT5P-110 + CuBERT & rrf (sum) & 100 & 99 & 97 & 88 & 36 & 2 & 70.33 & 68.66 & 1.67 \\
CT5P-110 + CuBERT & z-score (sum) & 100 & 99 & 97 & 86 & 34 & 1 & 69.50 & 68.66 & 0.84 \\
CT5P-110 + SPTCode & z-score (max) & 100 & 99 & 97 & 85 & 30 & 1 & 68.67 & 67.66 & 1.01 \\
CT5P-110 + SPTCode & rrf (max) & 100 & 99 & 97 & 85 & 30 & 1 & 68.67 & 67.66 & 1.01 \\
CT5P-110 + SPTCode & rrf (sum) & 100 & 99 & 97 & 85 & 30 & 1 & 68.67 & 67.66 & 1.01 \\
CT5P-110 + SPTCode & min-max (max) & 100 & 99 & 97 & 84 & 29 & 1 & 68.33 & 67.66 & 0.67 \\
CT5P-110 + SPTCode & min-max (sum) & 100 & 99 & 97 & 84 & 29 & 1 & 68.33 & 67.66 & 0.67 \\
CT5P-110 + SPTCode & z-score (sum) & 100 & 99 & 94 & 82 & 28 & 1 & 67.33 & 67.66 & \textit{-0.33} \\
\bottomrule
\end{tabular}

}
\end{table*}

\begin{table}[ht]
\renewcommand{\arraystretch}{1.5}
\centering
\caption{Precision for ensembles using different ensembling methods in In-situ-C-C++ dataset}
\label{tab:ensembles-in-situ}
\resizebox{0.5\textwidth}{!}{
\begin{tabular}{lrrr}
\toprule
\textbf{Ensembling Method} &
\textbf{CT5P-110} & \textbf{CuBERT} & \textbf{CuBERT} \\
& \textbf{+ CBFT} & \textbf{+ CBFT} & \textbf{+ CT5P-110} \\
\midrule
max\_individual       & 39.71 & 18.06 & 39.71 \\ \hline
min-max\_average      & 44.75 & 18.35 & 40.00 \\
min-max\_max          & 45.97 & 22.09 & 43.74 \\
min-max\_sum          & \textbf{46.91} & 22.37 & \textbf{44.39} \\
z-score\_average      & 42.01 & 20.86 & 35.97 \\
z-score\_max          & 43.67 & 24.24 & 39.78 \\
z-score\_sum          & 44.53 & \textbf{24.82} & 40.00 \\
rrf\_average          & 39.28 & 23.53 & 38.85 \\
rrf\_max              & 39.78 & 23.96 & 39.57 \\
rrf\_sum              & 39.71 & 24.17 & 39.78 \\ 
non-norm\_average     & 11.44 & 10.94 & 13.88 \\
non-norm\_max         & 16.04 & 17.19 & 21.15 \\
non-norm\_sum         & 19.50 & 19.64 & 23.24 \\ \hline
Absolute improvement (\%) & 7.2 & 6.76 & 4.68 \\
vs max\_individual &&& \\ \hline
Relative improvement (\%) & 18.13 & 37.43 & 11.79 \\
vs max\_individual &&& \\
\bottomrule
\end{tabular}}
\end{table}

\begin{table*}[h]
\renewcommand{\arraystretch}{1.5}
\centering
\caption{Statistical comparison of avg\_norm vs avg\_ind across different datasets.}
\resizebox{\textwidth}{!}{\begin{tabular}{lcccccccc}
\toprule
\textbf{Dataset (ensembling methods)} &\textbf{Metric} & \textbf{Sample} & \textbf{Avg.} & \textbf{Avg.} & \textbf{Diff} & \textbf{Normality} & \textbf{Test Used} & \textbf{p-value} \\
&&\textbf{Size}&\textbf{Ensembling} &\textbf{Ind. Max}&&\textbf{(p-value)}&& \\
\midrule
Company-C (all) & Avg. Recall&36 & 53.20 & 58.12 & -4.92 & 0.98621 & Paired t-test & \textbf{$< 0.00001$} \\
Company-C (excl. non-norm/average) & Avg. Recall&36 & 54.93 & 58.13 & -3.20 & 0.23017 & Paired t-test & \textbf{$< 0.00001$} \\
Company-CPP (all) & Avg. Recall&36 & 54.78 & 58.33 & -3.56 & 0.00428 & Wilcoxon test & \textbf{$< 0.00001$} \\
Company-CPP (excl. non-norm/average) & Avg. Recall&36 & 56.79 & 58.33 & -1.54 & 0.00396 & Wilcoxon test & \textbf{$< 0.00001$} \\
BCB13 & Recall&18 & 69.96 & 69.05 & 0.91 & 0.01409 & Wilcoxon test & \textbf{0.00004} \\
In-Situ-C-C++ (all) & Precision&36 & 30.73 & 32.49 & -1.76 & 0.00000 & Wilcoxon test & 0.44137 \\
In-Situ-C-C++ (excl. non-norm/average) & Precision&18 & 36.08 & 32.49 & 3.59 & 0.00610 & Wilcoxon test & \textbf{0.00059} \\
\bottomrule
\end{tabular}}
\label{tab:stat_tests}
\end{table*}

\subsubsection{Company-C and Company-C++ Results}

Tables~\ref{tab:ensembles-combinations-c} and ~\ref{tab:ensembles-combinations-cpp} summarize the average recall of each pairwise ensemble of models on Company-C and Company-C++ datasets, respectively, ranked by the \textit{Avg. Recall of ensembling methods}. For each ensemble in these tables, the average recall is across all cutoffs (4) and all possible ensembling methods (12). In turn, the average individual maximum recall is calculated across all cutoffs (4) for that ensemble. Likewise, \textit{Best Ensembling Recall} is the average recall of an ensembling method across cutoffs (4). Finally, the \textit{Best Ensembling Method} shows which method produced the best ensembling recall for that particular ensemble.

For Company-C (Table~\ref{tab:ensembles-combinations-c}), for three out of 36 combinations (highlighted in bold) a higher average recall (\textit{Avg. Recall of ensembling methods}) was observed in comparison to the larger of the two individual models (\textit{Avg. Ind. Max Recall}). For instance, combining \texttt{GCB} and \texttt{CBFT} yields $52.62\%$ average recall, narrowly exceeding the \texttt{max\_individual} recall of $52.14\%$ by $+0.48\%$. In another example (\texttt{StarEncoder + GCB}), the difference is $+1.34\%$.  The combination \texttt{C4 + CT5P-220} shows the largest positive difference ($+3.28\%$). However, when comparing \textit{Best Ensembling Recall} with \textit{Avg. Ind Max Recall} the former is better in 13/36 cases and in another 9/36 cases both are the same. This suggests that certain ensembling methods can be better than others: indeed, looking at \textit{Best Ensembling Method} in Table~\ref{tab:ensembles-combinations-c} ther are not average aggregations, for example.

A very similar pattern arises in Company-C++ (Table~\ref{tab:ensembles-combinations-cpp}). A few pairwise ensembles offer minor gains over the better of the two models (e.g., \texttt{SPTCode + CuBERT} at $+0.35\%$, \texttt{GCB + CuBERT} at $+0.32\%$, and \texttt{StarEncoder + GCB} at $+0.45\%$). In the majority of combinations, however, the averaged ensemble recall is slightly lower. Best-performing ensembles were better in 17/36 cases (highlighted bold) and were the same in 15/36 cases. Again, interestingly no average aggregation was observed among the best performing ensembling methods.

Looking at best-performing ensembles in Tables~\ref{tab:ensembles-combinations-c} and ~\ref{tab:ensembles-combinations-cpp} it can be seen that recall has improved in 30/72 cases (41.66\% of all cases), was similar in 24/72 cases (33.33\% of all cases) and worse in 18/72 cases (25\% of all cases).

\texttt{CT5P-110 + SPTCode}, \texttt{CT5P-110 + CuBERT}, and \texttt{CT5P-110 + CT5} appear to be the top 3 best-performing ensembles across these two datasets. Unsurprisingly, the individual models in these ensembles are also among the individual top-performing models, studied in RQ1 (see Table~\ref{tab:ranking_models}). 

Tables~\ref{tab:ensembles-c} and~\ref{tab:ensembles-cpp} report the average recall per \emph{ensembling method}, comparing them against the baseline \texttt{max\_individual} recall.  
As can be seen, these tables show that \texttt{max\_individual} yields the highest average recall in both Company-C ($58.63\%$) and Company-C++ ($58.83\%$). For Company-C, the best performing methods (excluding the \texttt{max\_individual} baseline) are \texttt{min-max\_max} ($56.77\%$) and \texttt{rrf\_max} ($55.55\%$); for Company-C++, top methods include \texttt{rrf\_sum} ($57.26\%$) and \texttt{z-score\_max} ($57.20\%$). At the same time, certain approaches such as \texttt{non-norm\_average} yield significantly lower recall in both datasets. This can also be seen in Figure~\ref{fig:heatmap1} and ~\ref{fig:heatmap2}, where \texttt{non-norm} normalization method and \texttt{average} aggregation method show systematically worse results than other. 

Overall, in both the Company-C and Company-C++ datasets, while certain ensembles sometimes give \emph{small} improvements, most pairs fail to surpass the better single model. Ensembling methods including \texttt{non-norm} normalization method and \texttt{average} aggregation method show consistently worse results. Finally, the best-performing ensembles seem to be constructed of likewise the best-performing individual models, like \texttt{CT5P-110 + SPTCode}.

\subsubsection{BCB13 Results}
For BCB13 evaluation, we selected the 3 top-performing ensembles from previous evaluation on Company-C and Company-C++ datasets: \texttt{CT5P-110 + SPTCode}, \texttt{CT5P-110 + CuBERT}, and \texttt{CT5P-110 + CT5}. We also excluded the worst performing ensembling methods containing either \texttt{non-norm} normalization method and \texttt{average} aggregation method, resulting in 18 ensemble configurations trialed (3 ensembles x 3 normalization methods x 2 aggregation methods).

Table~\ref{tab:ensemble_results_bcb} shows selected ensembles on BCB13 against their recall per clone Type, the average recall across clone types (Avg Recall), and maximum individual model’s recall (Ind Max). In contrast to Company-C and Company-C++, we observed that the majority of ensemble approaches exhibit a \emph{positive} difference over \texttt{Ind Max}. For example, the combination \texttt{CT5P-110 + CuBERT} with \texttt{min-max(max)} or \texttt{z-score(max)} outperforms its \texttt{max\_individual} by $+1.67\%$, and \texttt{CT5P-110 + SPTCode} gains roughly $+1\%$. Interestingly, the only two cases where the difference was negative, both used \texttt{z-score (sum)} ensembling method (highlighted in italics in the table). Therefore, ensembling on BCB13 consistently brings modest recall improvements avoiding that ensembling strategy. 

\subsubsection{In-Situ-C-C++ Results}
For in-situ evaluation, we used ensembles of models previously selected for individual evaluation (see Table~\ref{tab:true_clones_precision}) and all possible ensembling methods.

Unlike the recall-focused evaluations for Company-C, Company-C++, and BCB13, here we calculated \emph{precision} for In-Situ-C-C++ dataset as shown in Table~\ref{tab:ensembles-in-situ}. This was because of the nature of our evaluation, where a software engineer from the company manually evaluated each clone-pair candidate as a true positive or a false positive. But it should be noted that, as the first 1390 candidate pairs were assessed in each instance, the precision score also directly reflects the relative recall of the approaches over that 1390 dataset. Here, ensemble methods often outperform the stronger single model by a tangible margin. For instance, \texttt{CT5P-110 + CBFT} shows that several methods (\texttt{min-max\_sum}, \texttt{z-score\_sum}) achieve precision above $44\%$, whereas the stronger single model has $39.71\%$.  Similarly, \texttt{CuBERT + CBFT} experiences a marked jump from $18.06\%$ up to $24.82\%$ with \texttt{z-score\_sum}, and \texttt{CuBERT + CT5P-110} sees improvements of $+4.68\%$ with \texttt{min-max\_sum}. However, in line with our previous observations from the Company-C and Company-C++ datasets, inclusion of the \texttt{non-norm} normalization method and \texttt{average} aggregation method significantly reduces ensemble's recall. This is particularly pronounced for \texttt{non-norm} normalization method, where all ensembles trialed with this method reported significantly lower recall as compared to \texttt{max\_individual}. The inclusion of \texttt{average} also resulted in lower recall than if \texttt{max} or \texttt{sum} aggregation method was used.

\subsubsection{Statistical significance of ensembles}
The effects of ensembling were tested for their statistical significance as shown in Table~\ref{tab:stat_tests}. First, the distribution of sample data was checked for normality and the appropriate statistical test was used: paired t-test for normal distribution and Wilcoxon test for non-normal distribution \cite{chochlov2017historical}. 

As can be seen from the table, for Company-C/C++ datasets with all ensembling methods considered, ensembling is significantly worse than using the stronger individual model alone (p-value $< 0.0001$). If we exclude the worst-performing \texttt{non-norm\_average} methods, ensembling narrows the performance gap, yet still remains statistically lower for both Company-C and Company-C++.

Contrary, for BCB13 dataset (where \texttt{non-norm\_average} methods were excluded as discussed above) statistical testing shows that ensebmling improvement although moderate is still statistically significant($p=0.00004$).

Finally for the In-Situ-C-C++ dataset, including all methods yields an insignificant ($p=0.441$) negative precision difference, due to the poor performance of \texttt{non-norm\_average}. Excluding those methods, however, ensembling significantly outperforms the individual model’s precision by $+3.59\%$ on average ($p=0.00059$).  

\subsubsection{Answer to RQ3}
\label{subsec:rq3_discussion}

\paragraph{RQ3: How effective are ensembles of these LLMs?} 
Our results seem to demonstrate that the effectiveness of ensembling depends strongly on the ensembling methods and a dataset. For smaller Company-C and Company-C++ datasets, combining multiple LLMs frequently underperforms the single best model, and those differences are statistically significant. In contrast, for much larger BCB13 dataset, ensembles consistently outperform the stronger individual model by a moderate yet statistically significant margin. Finally, in the In-Situ-C-C++ scenario (precision-based), ensembling yields marked statistically significant improvements over individual models for when \texttt{non-norm\_average} methods are excluded achieving up to 7.2\% absolute and 37.43\% relative improvement for some best-performing ensembles over individual models (see Table~\ref{tab:ensembles-in-situ}).  

\paragraph{RQ3a: How do ensembling methods affect effectiveness?}
Certain normalization and aggregation strategies (\texttt{min-max\_sum}, \texttt{z-score\_sum}, \texttt{rrf\_sum}, \texttt{min-max\_max}, etc.) emerge as better-performing approaches across datasets (see Tables~\ref{tab:ensembles-c}, ~\ref{tab:ensembles-cpp}, ~\ref{tab:ensemble_results_bcb}, and ~\ref{tab:ensembles-in-situ}). On the other hand, methods such as \texttt{non-norm\_average} provide considerably lower performance, often dragging down ensemble results. This pattern highlights the importance of properly normalizing and aggregating the models’ similarity or rank scores before merging them. Even for those datasets where ensembling underperforms the single best model on average (Company-C and Company-C++), certain \emph{pairs} or specific \emph{methods} do lead to small gains. Meanwhile, for BCB-13 and In-Situ-C-C++, more widespread and substantial improvements are observed when the ensemble is formed using normalization and aggregation methods that exclude  \texttt{non-norm\_average}.

\subsubsection{Discussion}

\begin{itemize}
    \item \textbf{Performance Propagation in Ensembles}: the performance of an ensemble seems to be directly influenced by the strength of its constituent models. Indeed, the best performing individual models (e.g. CT5P-110, CuBERT, SPTCode and CT5 (see Table~\ref{tab:ranking_models})) are also constituent parts of the best performing ensembles across all datasets studied in this RQ. A key factor in any ensemble is the quality of its constituent models. Since each model contributes its own ranked list of candidate clones, stronger models (i.e., models that individually exhibit higher recall or precision) are more likely to contribute relevant clone candidates to the final merged list. By contrast, weaker models tend to introduce more noise or irrelevant candidates, which can obscure contributions from the stronger model: Even robust normalization and aggregation techniques cannot fully compensate for poor input quality. Thus, ensembling two powerful models has a higher chance of merging complementary strengths to yield an improvement, whereas combining a strong model with a markedly weaker one often does not surpass the stronger model by itself (for example, see CT5P-110 + C4 in Tables~\ref{tab:ensembles-combinations-c} and ~\ref{tab:ensembles-combinations-cpp}, where CT5P-110 is one of the best performing individual models, whereas C4 is one of the worst).
    \item \textbf{The Impact of Ensembling Methods}: when no normalization is used, each LLM’s raw similarity scores may lie on very different scales. Additionally, averaging these raw scores can effectively dilute or skew the impact of better-scoring candidates. For example, a weaker model whose scores are systematically higher or lower than those of a stronger model can dominate or under-represent the final ensemble scores in an uncalibrated way. Normalization first re-scales each model’s scores into a comparable range or based on rank (e.g., \texttt{min-max}, \texttt{z-score}, or \texttt{rrf}), preventing one model’s raw values from unfairly overshadowing or being overshadowed by another’s. Methods like \texttt{sum} and \texttt{max}, particularly when coupled with normalization, retain higher contributions from each model more effectively than simply taking an un-normalized average, thereby producing better overall performance.

    In this regard, some aggregation-normalization combinations show persistently strong results, particularly on company-provided datasets (both open and in-situ). Notably, \texttt{min-max-sum} and \texttt{min-max-max} perform consistently well, suggesting a possible C/C++ or company-specific bias. The effectiveness of \texttt{min-max-average}, although slightly lower, reinforces the importance of the \texttt{min-max} normalization step in aligning score distributions. Interestingly, \texttt{z-score-max} also achieves strong results on the company C/C++ data, though it is marginally outperformed by \texttt{z-score-sum}—a combination that performs poorly on other evaluation datasets. These trends highlight that some ensembling strategies may be more domain-sensitive than others, and further suggest that model fusion choices should consider both performance consistency and deployment context.
    
    \item \textbf{The Effect of a Dataset Size}: it was observed that ensembles do significantly worse on much smaller datasets like Company-C/C++, but significantly better on much larger datasets like BCB13 and In-Situ-C-C++ (see Table~\ref{tab:stat_tests}). One plausible explanation to this is that on smaller datasets, there is typically less diversity among clone candidates and fewer opportunities for models to complement each other, causing ensemble methods to contribute more noise than synergy. In contrast, larger datasets often present a broader range of coding patterns, giving each model more room to offer complementary coverage. Consequently, with larger code-bases for investigation, the benefits of combining multiple perspectives in an ensemble can outweigh any added noise or overlap, leading to stronger overall performance.
    \item \textbf{The Implication for Computational Resources}: while ensembling seems to be significantly advantageous on larger datasets, its application comes at a price of increased hardware/time resources needed. Particularly, based on the ensembling approach used, it could be either memory-intensive (if parallelization of ensembles is employed) or time-intensive (if consecutive ensembling is employed). In our trials with the larger BCB13 dataset on the M1 machine, time-demanding consecutive ensembling was employed. For example, the execution of CT5P-110 model resulted in \textit{1h33m34s} of inference time (we omit the parsing time here because it is constant for the same dataset and we omit the search time because it is insignificantly small as compared to inference \cite{chochlov2022using}. Likewise ensembling time is small and is omitted as well) and the execution of the CT5 model took \textit{1h30m17s}. Therefore, the ensembling time here becomes \textit{3h03m51s}, almost doubling that of individual models.
\end{itemize}

\section{Threats to validity}
\label{sec:threats}
In this section, we discuss potential limitations and biases in our study, following common guidelines for empirical software engineering \cite{kitchenham2004evidence}. We group threats into four categories: construct, internal, external, and conclusion validity.

\subsection{Construct Validity}
In this study, we used \emph{recall} and to a lesser extent \emph{precision} metrics as indicators of LLM effectiveness. These metrics are widely accepted in clone detection research, but are not exhaustive and other metrics like accuracy could have been applied. Other potential construct-validity issues include:
\begin{itemize}
    \item \textbf{BCB13 evaluations using recall only:} Only recall can be computed automatically in BCB13. Although widely practiced, omitting precision might limit the overall picture of performance: 100\% recall can be obtained by returning every possible combination of code-pairs - a strategy disallowed by having a suitable precision threshold. Manual sampling of candidates is possible but can be subjective \cite{Li2020, chochlov2022using}. Instead, to mitigate this issue, a common \texttt{global Top K} cutoff threshold was applied across all models. The rationale behind this is that, if LLM A recalls more clone candidates within the same cutoff threshold than LLM B, then LLM A is likely to have better precision, assuming other factors are constant. This is because the precision metric measures the proportion of true positive clones among the detected candidates, and having a higher number of candidates within the same threshold suggests better selection or ranking of relevant clones.
    \item \textbf{In-situ evaluations relying on precision only:} For the In-Situ-C-C++ dataset, we relied on expert assessments of true and false positives, allowing for the calculation of precision. It was impossible to obtain recall results because we did not know all the clones in the dataset, and so could not determine the false negatives. However, because all the results were again calculated based on the same number of clone-pair candidates proposed by the approach (1390), the precision results obtained were a proxy for the number of clones identified, which was useful in comparing across the approached employed. Differences in expert judgment in determining true positives and false positives may also have arisen, but the engineer involved followed strongly-defined company guidelines to support their judgment.
    %\item \textbf{Similarity scores and thresholds:} Each LLM produces raw similarity scores on different scales. Normalization methods and thresholds used (e.g., \texttt{z-score}, \texttt{min-max}) may affect which clones are ranked higher and thus declared “true” or “false” by domain experts.
\end{itemize}

\subsection{Internal Validity}

\begin{itemize}
    \item \textbf{Selection of LLMs:} We chose 9 transformer-based LLMs after filtering a larger pool. While our criteria (public availability, source-code pretraining, etc., see Section~\ref{sec:methodology}) were designed for reproducibility, excluding proprietary or partially documented models could introduce selection bias.
    \item \textbf{Scalable batch search approach:} We used SSCD with parameters tailored to each dataset (e.g., \texttt{top N clone class candidates}, \texttt{global top K}). Different parameter choices could alter the set of candidates retrieved, potentially affecting recalls.
    \item \textbf{Threshold tuning in in-situ evaluation:} For the private dataset, thresholds were chosen partially by prior experience and pilot testing. This tuning could favor certain models or artificially inflate (or deflate) their measured precision.
    \item \textbf{Confounding factors in regression analysis:} The OLS and Elastic Net regressions (RQ2) might omit relevant variables (e.g., specific hyperparameters, training schedules). Although we tested for multicollinearity and used multiple models, unmeasured confounders may remain.
\end{itemize}

\subsection{External Validity}
\begin{itemize}
    \item \textbf{Dataset diversity:} We covered varying dataset sizes (from tens of thousands to millions of LOC) and three popular languages (C/C++/Java). Although these represent multiple real-world scenarios, the findings may not fully generalize to other programming languages (e.g., Python, JavaScript) or domain-specific codebases.
    \item \textbf{Industrial partner context:} The In-Situ-C-C++ dataset captures only one company’s code, coding styles, and domain constraints. Results in different industrial settings (with different code organization or development processes) might vary.
    \item \textbf{Limited scope of ensembling:} We explored only pairwise ensembles of 9 LLMs and tailored this number further for the BCB13/In-Situ-C-C++ datasets selecting only top-performing models/ensembles. Combining more than two models or including additional model families could yield different outcomes, potentially influencing both recall and precision.
\end{itemize}

\subsection{Conclusion Validity}
\begin{itemize}
    \item \textbf{Statistical tests:} We used paired t-tests and Wilcoxon tests depending on normality checks. Although appropriate, the number of samples in some comparisons (e.g., 36 pairs, 18 pairs, etc.) may still limit power in detecting smaller effect sizes.
    \item \textbf{Sample size in regression (RQ2):} With only 27 total observations (9 LLMs $\times$ 3 datasets), the regression models risk overfitting. We partially mitigated this with OLS and Elastic Net cross-validation, but the limited sample can reduce generalizability of coefficient estimates.
    \item \textbf{Variability in clone labeling:} Even though we used established datasets (Company-C, Company-C++, BCB13) and an expert-based approach for In-Situ-C-C++, minor inconsistencies in labeling or ground truth definitions of clones could affect measured recall and precision values.
\end{itemize}

\section{Conclusions and future work}
\label{sec:conclusions}

This paper presented an empirical study of 9 transformer-based LLMs for scalable code clone detection. Our research explored three key aspects:

\begin{itemize}
    \item \textbf{RQ1: How effective are novel LLMs for scalable clone detection?}  
    Experimental results across three public datasets (Company-C, Company-C++, and BCB13) and a private industrial dataset (In-Situ-C-C++) suggest that, while no single LLM \emph{uniformly} excels, certain models like CodeT5+ 110M (CT5P-110) and CuBERT emerge as top performers. Interestingly, these rankings are dataset-dependent, highlighting that factors such as clone class size can significantly affect each model’s recall. In an in-situ industrial evaluation, CT5P-110 notably achieved high precision (39.71\%), significantly outperforming existing baselines like fine-tuned CodeBERT (CBFT) and CuBERT, which underscores the \emph{practical utility} of carefully selecting LLMs in real-world settings.

    \item \textbf{RQ2: How do characteristics of these LLMs affect their effectiveness with regard to recall?}  
    Regression analysis (using both OLS and Elastic Net) reveals that \emph{training data composition} and \emph{tokenizer configurations} have statistically significant impacts on recall. Counter to expectations, larger embedding sizes and vocabulary sizes correlate with \emph{lower} recall, while the number of encoder parameters (i.e., model size) is not a statistically significant predictor. Models trained on datasets beyond CodeSearchNet tend to fare better, suggesting that data diversity or domain alignment may be more critical than sheer model size. This aligns with LLM development trends, suggesting a more \textit{compact} LLM can be advantageous for large scale clone detection with more focus dedicated to training data and its quality \cite{roziere2023code}.

    \item \textbf{RQ3: How effective are ensembles of these LLMs, and RQ3a: how do ensembling methods affect effectiveness?}  
    Our findings show that ensembles can substantially improve detection performance—particularly on \emph{larger} datasets (BCB13) and in a real-world in-situ scenario. However, for smaller datasets such as Company-C and Company-C++, ensembles can underperform when compared to the single best model. Effective normalization (\texttt{z-score}, \texttt{min-max}, \texttt{rrf}) and aggregation (\texttt{sum}, \texttt{max}) are essential; the \texttt{non-norm\_average} method consistently leads to reduced performance. In sum, the impact of ensembling is \emph{context-sensitive}, benefiting most when data diversity is high and constituent models have complementary strengths.
\end{itemize}

Overall, these findings indicate that \textbf{LLMs and ensembles of LLMs can be very effective for code clone detection at scale, even if it can still be difficult to identify the specific LLMs and ensembles that should be employed}. %but success depends on \emph{selecting models with complementary training data, carefully tuning tokenizer configurations, and using appropriate ensembling strategies when beneficial}. 

Several future work directions can arise from the insights of this study:

\begin{enumerate}
    %\item \textbf{Inclusion of Additional Model Families and Larger Parameter Variants:}  
    %Although we explored nine transformer-based models, other emerging architectures (e.g., hybrid transformer-graph neural networks, retrieval-augmented transformers) could enrich the comparison further. Additionally, evaluating significantly larger parameter variants (e.g., 7B, 13B, and beyond) under the same experimental framework would help confirm or refute the minimal impact of model size on clone detection.

    \item \textbf{Extended Datasets and Programming Languages:}  
    Our focus on C/C++ and Java provided valuable insights, but real-world systems often have other languages like Python, JavaScript, Go, and domain-specific languages. Future work can assess whether the observed patterns (e.g., irrelevance of the number of parameters) and findings persist in broader, multilingual contexts.

    \item \textbf{Ensembling More Than Two Models:}  
    We limited our study to pairwise ensembles for manageability and interpretability. Investigating multi-model ensembles can suggest how effects change, albeit at the cost of computational resources and increased complexity in analysis.

    \item \textbf{Refined Normalization and Aggregation Approaches:}  
    Our findings highlight the importance and variability of carefully calibrated similarity scores and suitable aggregation (e.g., \texttt{sum}, \texttt{max}). Future work might adapt advanced ranking-fusion techniques or dynamic weighting schemes (e.g., learning to rank) \cite{liu2009learning} to optimize ensemble performance on heterogeneous datasets.

    \item \textbf{Exploration of Additional Metrics and Real-World Constraints:}  
    Although recall and precision remain fundamental, industrial practitioners also care about overall impact in terms of costs, runtime overhead, and interpretability of clone detection outputs. Follow-up studies could incorporate these aspects, aiming to propose comprehensive metrics that capture real-world utility beyond pure detection rates.

    %\item \textbf{Longitudinal / Evolutionary Studies:}  
    %Most modern codebases evolve rapidly. Investigating how LLM-based clone detection adapts to continuous integration and frequent code changes would be an important extension. Tracking precision and recall across multiple software releases could highlight challenges like model drift or changes in token distributions.

\end{enumerate}

\begin{acks}
This work was supported, in part, by Science Foundation Ireland grant 13/RC/2094\_2 and by the participating company.
\end{acks}

\bibliographystyle{ACM-Reference-Format}
\bibliography{refs}

\end{document}